\documentclass[12pt]{elsarticle}

\usepackage[T1]{fontenc}
\usepackage{graphicx}
\usepackage{xcolor}
\usepackage{dcolumn}
\usepackage{booktabs}
\usepackage{bm}
\usepackage[nice]{nicefrac}
\usepackage{amsmath}
\usepackage{amssymb}
\usepackage{scalerel}
\usepackage{array}
\usepackage{hyperref}
\usepackage{cleveref}

\newcolumntype{C}[1]{>{\centering\arraybackslash}p{#1}}

\definecolor{blush}{rgb}{0.87, 0.36, 0.51}
\newcommand{\eq}[1]{Eq.~\ref{#1}}
\newcommand{\Eq}[1]{Equation~\ref{#1}}
\newcommand{\fig}[1]{Fig.~\ref{#1}}

\newcommand{\red}[1]{{\color{blush}#1}}

\newcommand{\blue}[1]{#1}
\newcommand{\E}{\mathcal{E}}
\newcommand{\scale}[1]{\scaleto{#1}{5pt}}

\newcommand{\ms}{\,\mathrm{m/s}}

\newcommand{\orcid}[1]{
\href{https://orcid.org/#1}{\includegraphics[width=10pt]{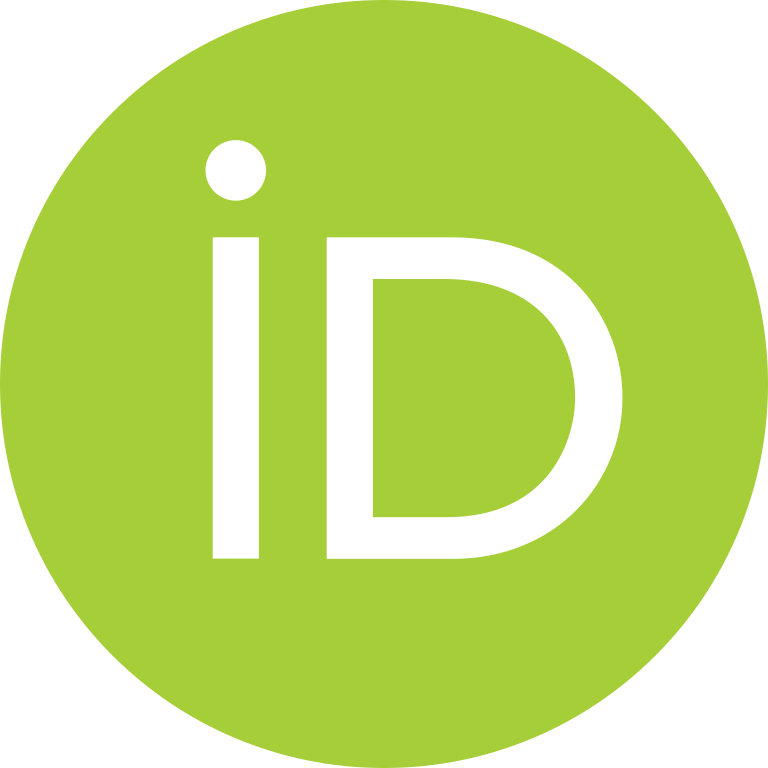}}}

\newcommand{\rev}[2]{#2}

\usepackage{times}

\hypersetup{
    colorlinks,
    linkcolor={blush},
    citecolor={blue!70!black},
    urlcolor={blue!70!black}
}

\begin{document} 

\begin{frontmatter}

\title{Body and mind: Decoding the dynamics of pedestrians and 
the effect of smartphone distraction
by coupling mechanical and decisional processes} 

\author[CFTC,UNAV]{I\~naki ECHEVERR\'IA-HUARTE  \corref{cor} \orcid{0000-0001-6992-0815} }
\ead{iecheverria.13@alumni.unav.es}


\affiliation[CFTC]{organization={Centro de F\'{i}sica Te\'{o}rica e Computacional, Faculdade de Ci\^{e}ncias, Universidade de Lisboa},
            city={Lisboa},
            postcode={1749-016},
             country={Portugal}}

\affiliation[UNAV]{organization={Departamento de F{\'{i}}sica y Matem{\'{a}}tica Aplicada, Facultad de Ciencias, Universidad de Navarra},
            city={Pamplona},
            postcode={31080},
             country={Spain}}

\author[ILM]{Alexandre NICOLAS \corref{cor} \orcid{0000-0002-8953-3924}}
\ead{alexandre.nicolas@cnrs.fr}

 \affiliation[ILM]{organization={Institut Lumi{\`e}re Mati{\`e}re, CNRS, Univ. Claude Bernard Lyon 1},
            city={Villeurbanne},
            postcode={F-69622},
             country={France}}

\cortext[cor]{Corresponding authors}


\begin{abstract}
Pedestrians are able to anticipate, which gives them an edge in avoiding collisions  and navigating in cluttered spaces. However, these capabilities are impaired by digital distraction through smartphones, a growing safety concern. To capture these features, we put forward a continuous agent-based model (dubbed ANDA) hinging on a transparent delineation of a decision-making process,  wherein a desired velocity is selected as the optimum of a perceived cost, and a mechanical layer that handles contacts and collisions. Altogether, the model includes less than a dozen parameters, many of which are fit using independent experimental data.
The versatility of ANDA is demonstrated by numerical simulations that successfully replicate empirical observations in a very wide range of scenarios. These scenarios vary from collision avoidance involving one, two, or more agents, to collective flow properties in unidirectional and bidirectional settings, and to the dynamics of evacuation through a bottleneck, where contact forces are directly accessible. Remarkably, the model was able to replicate the enhanced chaoticity of the flow observed experimentally in 'smartphone-walking' pedestrians, by reducing the frequency of decisional updates, replicating the digital distraction effect. The conceptual transparency of the model makes it easy to pinpoint the origin of its current limitations and to clarify the singular position of pedestrian crowds amid active-matter systems.
\end{abstract}


\end{frontmatter}


\section{Introduction}

Pedestrians routinely display remarkable navigation and coordination abilities, which enable them to adapt to new environments, make their way through dense crowds \cite{bruneau2015energy,nicolas2019mechanical} and navigate in very constrained surroundings. 
But, just like Marcus Aurelius's infallible man \cite{aurelius2013marcus},
the infallible pedestrian simply does not exist: Suboptimal routing choices  \cite{gabbana2022fluctuations}, collisions, or even in the most tragic cases stampedes \cite{helbing2012crowd} are indeed also prominent features of crowd dynamics. Uncoordinated behavior gets even more visible
in our overly connected societies, where the pedestrians' attention to their surroundings is often diverted by their smartphones \cite{murakami2021mutual,jiang2018effects}.
Crowds thus display both high abilities for self-organization \emph{and} individualistic choices conducive to undesirable collective effects. Models capable of capturing this somewhat contradictory alliance would be highly beneficial for practical purposes, of course, when it comes to designing new pedestrian facilities \cite{alaska2017impact}, but also more fundamentally, to disentangle the specifics of pedestrian dynamics from the roots they share with other physical assemblies, notably active matter. At present, these antagonistic features are, to say the least, only dimly reflected in the vast array of microscopic models for crowd dynamics. Schematically, one branch of models prohibits the selection of all velocities potentially leading to a collision, whereas a second branch handles collision avoidance mechanistically, as a repulsive force.

The first branch (typically comprised of velocity-based models) was largely inspired by the field of robotics \cite{van2008interactive,curtis2013pedestrian,bareiss2015generalized}
in an endeavor to guarantee collision-free motion of multiple agents \cite{godoy2016c}. If it is implemented in a fully decentralized way, this approach tends to be overly conservative (`prudent') and too often the dynamics get frozen (deadlocks) or look unnatural in the presence of conflicting maneuvers \cite{shiomi2014towards}. To circumvent this issue, global coordination of individual moves may be enforced via a more or less centralized process \cite{godoy2016c,karamouzas2017implicit}. This leads to reasonable output for a variety of situations, but may arguably not be scalable to large crowds (involving thousands of pedestrians). Furthermore, the predicted trajectories 
\rev{tend to look too `robotic'}
{may look odd for pedestrians (with sharp turns, grazing trajectories, etc.)} \cite{curtis2013pedestrian,shiomi2014towards,karamouzas2017implicit}. The prediction of the other agents' trajectories, mediated by perception, can also be enforced in a context-dependent way \cite{wolinski2016warpdriver} and bring the agents' behavior closer to human response, but at the risk of requiring a different treatment for every situation and making them less amenable to theoretical understanding.

At the other extreme, in the wake of the celebrated Social Force Model \cite{helbing1995social}, force-based models 
hypothesize that the local rules  of navigation can be represented by \emph{ 
\rev{ad hoc}{specifically tailored}} 
pseudo-forces encoding `intuitive' social interactions (such as keeping some distance from one another via a repulsive or walking in a group via an attractive interaction \cite{zanlungo2014potential}) and inserted into an e.g. Newton-like equation of motion, along with mechanical forces. This particle-based approach has succeeded in replicating various collective and/or self-organized phenomena in crowds \cite{helbing2005self}, but is also known to lead to spurious oscillations \cite{maury2018crowds} and to deadlocks or conflicts caused by an unrealistic lack of anticipation by the agents.
These issues are partly remedied by supplementing the models 
\rev{with specific forces}{with specific forces or mechanisms}
enforcing 
a following or an anticipatory behavior \cite{shiomi2014towards,lu2020pedestrian}. In a study of note, Karamouzas et al. contended that collision avoidance is controlled by the anticipated time to collision (TTC) with somebody else, rather than the absolute distance $r$ to this person, and they showed that 
the distribution of spacings between moving pedestrians in
empirical datasets is better described using an interaction depending on the TTC, in lieu of $r$ \cite{Karamouzas2014universal}.
Other anticipatory behaviors and follow-the-leader rules have also recently been put forward \cite{shiomi2014towards,lu2020pedestrian},
but this remedial process 
\rev{is essentially \emph{ad hoc} and adds patches to an existing model without fundamentally questioning its overall structure.}{has fallen short of fundamentally questioning the overall structure of the model.}

\blue{
In short, agents of the first branch of models are too infallible in their will to avoid contacts, whereas the second branch underestimates the anticipation capabilities of pedestrians, notably in crowded conditions. This may explain why no model has succeeded yet
in quantitatively replicating very diverse empirical scenarios 
\rev{without specific adjustments of its rules or parameters.}
{without adjustments of its rules and/or specific calibration of its parameters for each class of scenarios.}
In this paper, we propound a modeling framework that restores the `fallibility' of pedestrians, who may accidentally bump into each other,
but generally find ways to navigate smoothly even in cluttered environments. It aims to
}
mirror the main processes involved in pedestrian motion, whereby each agent
updates their desired velocity via a decision-making process that optimizes a pseudo-energy (or perceived cost) \cite{moussaid2011simple}, notably comprising a
\rev{}{floor field that guides agents along the most suitable routes towards their target and a}
\rev{TTC term to render collision avoidance}
{mathematically smooth TTC penalty to avoid collisions all the more urgently as they are imminent} \cite{Karamouzas2014universal}, whereas the contacts and pushes 
that may ensue are handled by a mechanical layer (Sec.~\ref{sec:model}). Thanks to the transparent coincidence between the building blocks and the processes they describe, most model parameters can be calibrated independently. 
Inherently decentralized, our ANticipatory Dynamics Algorithm (ANDA) reproduces realistic collision avoidance in crowds 
and coordinated motion in crowded scenarios as well as other collective effects, in quantitative agreement with experimental data, using a single set of parameters for the different regimes under study (Sec.~\ref{sec:results}). The model can further straightforwardly be extended to account for `smartphone-walking', which has become a serious practical issue. The relative simplicity of the proposed framework makes it suitable for physical insight into the similarities and discrepancies with other types of active matter.

\section{\label{sec:model}Modeling framework}

\begin{figure*}[htp]
  	\centering
    \includegraphics[width=1\textwidth]{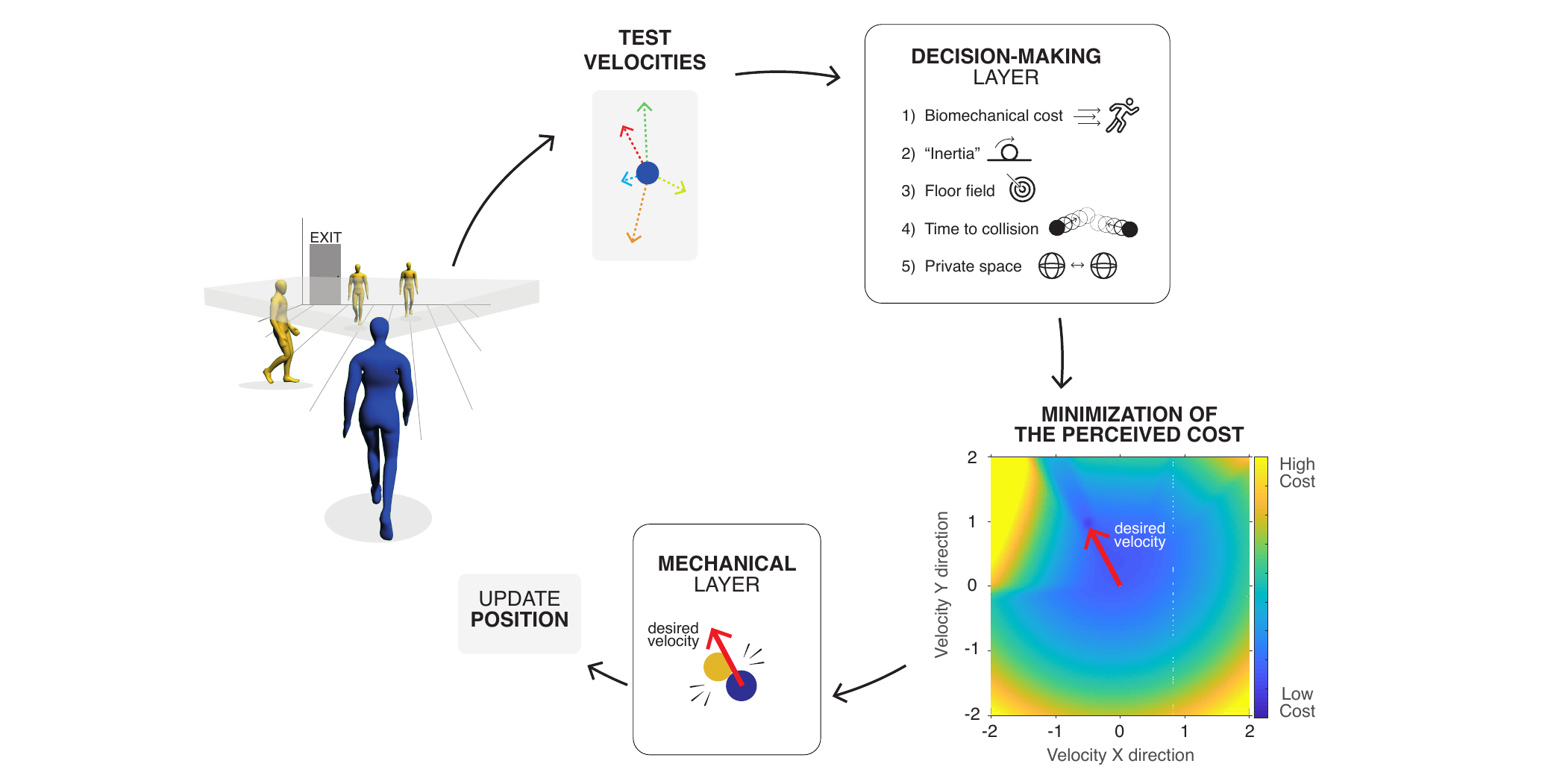}
    \caption{Schematic representation of the algorithm. Every time step $\delta t$, the agent collects information about the environment and the other agents in his or her field of view (left), considers various test velocities, and selects the optimal one 
    from the perspective of his/her perceived cost. The self-propulsion force corresponding to this desired velocity is implemented in the mechanical layer, which handles possible contacts or collisions and returns the agent's updated position.}
  	\label{fig:Sketch}
\end{figure*}

\subsection{Decision-making layer and mechanical layer}

A pedestrian is both an autonomous agent that controls his or her motion and a physical body that evolves in a mechanical environment. In the parlance of control theory, a pedestrian is thus \emph{both} the `controller' and the `system' responding to the control signal \cite{hoogendoorn2003simulation}. These two roles are amalgamated in force-based models for
pedestrian (and animal) motion \cite{maury2018crowds}, whose velocity is governed by a single equation.
Here, to mimic the sequential process at play in human locomotion, we choose to clearly disentangle the decision-making process \cite{hoogendoorn2003simulation,curtis2013pedestrian,moussaid2011simple,nicolas2020dense}, whereby the agent selects a desired velocity \rev{}{depending on various (biomechanical and psychological) factors}, from the mechanical block governing the response of the pedestrian's body in interaction with the environment.
\rev{}{These two blocks are handled sequentially for each agent, following the general scheme outlined in Fig.~\ref{fig:Sketch}, and will be exposed in detail below.}
  
\subsection{Decision-making layer}

The move (consciously or unconsciously) chosen by a pedestrian at each time step can be regarded as an optimum over a set of options from the viewpoint of this agent, i.e., the option that maximizes a utility or, equivalently, minimizes a perceived cost function $\E$ (regardless of whether this cost function is objective or subjective and assessed with exhaustive or limited information).
Such an approach has previously been applied for the selection of an optimal step \cite{antonini2006discrete}
\rev{}{or of an inter-pedestrian distance \cite{zanlungo2014potential}}, but here we apply it for the choice of a desired velocity $\boldsymbol{u^\star}$ for the next time step, viz.

\begin{equation}
\displaystyle
    \boldsymbol{u^\star} = \underset{\boldsymbol{u} \in \mathbb{R}^2}{\mathrm{arg min}}\ \E( \boldsymbol{u}).
\label{eq:argmin}
\end{equation}

From a broader perspective, this criterion can be interpreted as an optimal control problem \cite{hoogendoorn2003simulation} over a very small time horizon $\delta t$ (where $\delta t$ is the interval between decisions, due to the reaction time), in which one would like to extremize

\begin{equation*}
\centering
\displaystyle
\E[ \boldsymbol{u} ]  = 
   \underset{\text{running cost}}
   {\underbrace{\int_{t}^{t + \delta t}e\left(t',\boldsymbol{r}(t'),\boldsymbol{u}(t')\right) dt'}}
   +
   \underset{\text{terminal cost}}
   {\underbrace{\E^{\mathrm{T}}\left(\boldsymbol{r}(t+\delta t)\right)}}
\end{equation*}
\begin{equation}
  \E( \boldsymbol{u} )\overset{\delta t\to 0}{\approx} 
  \delta t\, e \left(t,\boldsymbol{r}(t),\boldsymbol{u}\right) +  \E^{\mathrm{T}}\left(\boldsymbol{r}(t)+\delta t\,\boldsymbol{u}\right)
 \label{eq:E_T}
\end{equation}
Equation~\ref{eq:argmin} is reminiscent of the least action principle for a physical system, whereby the trajectory selected by nature minimizes a quantity called the action (also see \cite{corbetta2019path} for an application to
dilute pedestrian flows). But, crucially, the cost is here minimized by each agent separately, knowing 
some information about the others, and not globally by the whole assembly, as in previous 
endeavors \cite{karamouzas2017implicit}. This reflects the autonomous nature of the agents and drives a wedge between social assemblies and physical systems.

\rev{Let us now detail the various terms contributing to the perceived cost $\E( \boldsymbol{u})$.}
{Specifying the decisional layer thus boils down to defining the perceived cost $\E( \boldsymbol{u})$. Given the complexity inherent in pedestrians' choices, $\E( \boldsymbol{u})$ will naturally comprise various terms, combined linearly here. Let us now detail these terms, each of which may become essential in some situations and reflects a specific motivation, either independently of the rest of the crowd (the eagerness to reach one's goal, the inconvenience of walking too fast or turning too sharply) or owing to social interactions (the reluctance for encroachments into personal spaces or collision paths). }

\subsubsection{Static floor field}
\label{sub:floor_field}
 In \eq{eq:E_T}, the driving term, which accounts for the (motivation-related) desire to move towards a target destination, is the terminal cost $\E^{\mathrm{T}}$, which is a function of the expected position $\boldsymbol{r'}$ at time $t+\delta t$:  The closer one gets to the target, the better. 
 For simplicity, we will assume that $\E^{\mathrm{T}}(\boldsymbol{r'})$ is proportional to the shortest-path distance $\mathcal{D}(\boldsymbol{r'})$ to the agent's target, defined by the Eikonal equation $|\nabla_{\boldsymbol{r}} \mathcal{D}|=n(\boldsymbol{r})$ (see \emph{Methods}). Here, the `refractive index'  $n(\boldsymbol{r})$ measures how uncomfortable the environment at $\boldsymbol{r}$ is. 
In particular, while $n=1$ in free space, proximity to a wall is penalized by the function
$n(\boldsymbol{r})=1 / \mathrm{tanh}(d_w(\boldsymbol{r})/d_c)$, where $d_w(\boldsymbol{r})$ denotes the
distance to the closest wall and the repulsive length $d_c$
is a parameter of the model. Overall, the floor field thus reads
\begin{equation}
    \E^{\mathrm{T}}(\boldsymbol{r'}) = \frac{K^T}{n}\, \mathcal{D}(\boldsymbol{r'}),
    \label{eq:ET_D}
\end{equation}
where $K^T>0$ is a coefficient which we have chosen to divide by $n=n(\boldsymbol{r})$, for reasons that will soon transpire (Sec.~\ref{sec:speed_energy}). 
Importantly, using a floor field in ANDA circumvents the practical issues arising from the prescription of standard road maps (i.e., `central' paths) or way-points in a complex geometry \cite{curtis2013pedestrian}, which 
\rev{is here handled straightforwardly}
{can here be handled in a straightforward and efficient way (see Sec.~\ref{sub:complex_geometry} and Appendix~C).}

\subsubsection{Bio-mechanical cost associated with walking speed and `inertia'}
\label{sec:speed_energy}
The target cannot be reached instantly: The locomotive abilities of pedestrians constrain the choice of a desired velocity $\boldsymbol{u^\star}$. 
The faster one moves, the more energy is consumed by the body. The bio-mechanical dependence of this energy $e^{\mathrm{speed}}$ on speed
$u=||\boldsymbol{u}||$, which is internalized in the decision-making process, has been quantitatively assessed via measurements of oxygen consumption (e.g., of participants walking on a treadmill) \cite{ludlow2016energy}; the excess energy expenditure compared to rest comprises a constant (penalty for walking) plus a term which
grows as the square of $u$ multiplied by a height-dependent prefactor. We show in \red{Appendix~A} that these experimental data 
are nicely fit by
\begin{equation}
\displaystyle
    e^{\mathrm{speed}}(u)=
    \begin{cases}
7.6\,u-35.4\,u^{2} & \text{ for } u <0.1\ms\\
0.4+0.6\,u^{2} & \text{ for }u\geqslant0.1\ms.
\end{cases}
    \label{eq:Espeed}
\end{equation}
which smoothly connects $e^{\mathrm{speed}}(u)$ to 0 when $u\to 0$, i.e., when an agent halts.

Besides, abrupt changes in velocity should also be barred because they are uncomfortable and bio-mechanically costly, which suggests an `inertial' contribution

\begin{equation}
\displaystyle
    e^{\mathrm{inertia}} (\boldsymbol{u})= \mu \Big(\boldsymbol{u} - \boldsymbol{v}(t)
    \Big)^2,
    \label{eq:EInertia}
\end{equation}
where $\boldsymbol{v}(t)$ is the actual velocity at time $t$ and $\boldsymbol{u}$, the test velocity for the next time step, $t+\delta t$, and $\mu>0$.

\rev{Without any further contributions(from the environment),}
{The terms $\E^{\mathrm{T}}(\boldsymbol{r'})$, $e^{\mathrm{speed}}(u)$, and $e^{\mathrm{inertia}}$ presented so far are the only ones
affecting an isolated pedestrian. In this situation,}
 the extremum of $\E(\boldsymbol{u})= \delta t \Big[e^{\mathrm{speed}}(\boldsymbol{u}) + e^{\mathrm{inertia}}(\boldsymbol{u}) \Big]+ \E^T(\boldsymbol{r}+\delta t\, \boldsymbol{u})$ is reached when
\begin{equation}
    0 = \frac{1}{\delta t}\nabla_{ \boldsymbol{u}} \E 
    = 2\, \mu \Big(\boldsymbol{u} - \boldsymbol{v}(t)   \Big) + \frac{d e^{\mathrm{speed}}}{du}   \frac{\boldsymbol{u}}{u} - K^T \, \boldsymbol{t},
\label{eq:free_vd}
\end{equation}
where $\boldsymbol{t}=-\nabla_{ \boldsymbol{r}} \mathcal{D}/n(\boldsymbol{r})$ is a unit vector pointing towards the target, in free space.  Provided that  the free walking speed $u^\infty$ \emph{in steady state} is empirically known, the parameter $K^T=1.2\,u^\infty$, for $u\geqslant 0.1\ms$, can be set straightforwardly; details can be found in \red{Appendix~B}. 
\rev{}{In passing, the foregoing equality, balancing motivation and biomechanical cost, also shows how being more eager to egress e.g. in the context of an evacuation (see Sec.~\ref{sub:evacuation}), as materialized by a steeper slope $K^T$ of the floor field, translates into a higher free walking speed. }

Strikingly, since $\boldsymbol{u}$ is the desired velocity at $t+\delta t$,  \eq{eq:free_vd} is formally identical to the numerical resolution of Newton's second law with a forward Euler scheme \cite{karamouzas2017implicit}. 
At this stage, we should underscore the two key conceptual shifts that have been made so far with respect to most existing models. First, Newton's equation is not obtained by dint of some fundamental physical law, but because of the simple form chosen for $e^{\mathrm{inertia}}$ in the decisional layer. Second, the free walking speed $u^\infty$ explicitly results from the balance between a bio-mechanical cost $e^{\mathrm{speed}}$, that may vary with the pedestrian but \emph{not} with the context, and a will to move described by the terminal cost (or floor field) $\E^T$; all psychological and motivational effects (heightened in the event of an emergency, for instance) are deferred to $\E^T$. This makes sense because $\E^T$ also governs route choice and it will prove instrumental in dealing with complex obstacles \red{(Appendix~C)} and geometries (Sec.~\ref{sub:complex_geometry}), 
which stand out as pitfalls for other models.

\subsubsection{Personal space}
Another intuitive contribution comes from the reluctance to stand excessively close to other pedestrians or obstacles, i.e., to preserve one's personal space, the size of which varies between cultures \cite{sorokowska2017preferred} 
\rev{.}{ and, more anecdotally, if social-distancing measures are in place \cite{pouw2020monitoring,echeverria2021effect}. This contribution is particularly important for dense static crowds \cite{nicolas2019mechanical,cordes2023dimensionless}. }
Here, it is modeled as a distance-dependent repulsive term entering the terminal cost in \eq{eq:E_T},

\begin{equation}
    \E^{\mathrm{personal}}(\boldsymbol{r'}) = 
    \sum_{j \in f.o.v.(i)} \frac{\eta}{\sigma_i + \sigma_j} V^{\mathrm{rep}}\Big(\frac{||\boldsymbol{r'} - \boldsymbol{r}_j(t+\delta t)||}{\sigma_i+\sigma_j}\Big),
\label{eq:E_rep}
\end{equation}
where $\boldsymbol{r}_j(t+\delta t)=\boldsymbol{r}_j(t)+\delta t\,\boldsymbol{v}_j(t)$ is the expected position of agent $j$ at the next time step,  $\sigma_j$ is the radius of agent $j$, and $V^{\mathrm{rep}}$ is a short-ranged function. We cut it off at $1+\epsilon^{\star}$, where $\epsilon^{\star}\geqslant 0$, and set

\begin{eqnarray}
V^{\mathrm{rep}}(\tilde{r})	=	
\begin{cases}
\frac{1}{\tilde{r}}-\frac{1}{1+\epsilon^{\star}}
& \text{ if } \tilde{r}<1+\epsilon^{\star}\\
0 & \text{ otherwise.}
\end{cases}
\end{eqnarray}

To account for perception, in \eq{eq:E_rep}, the sum does not run over all neighboring agents $j$, but is limited to those in the \emph{field of view} (f.o.v.) of the agent ($i$), i.e., within a cone which extends from $-\theta$ to $+\theta$ around the direction of the agent's last desired velocity $\boldsymbol{u}(t)$.

\subsubsection{Time-to-collision (TTC) energy}

All contributions so far are based on positions and distances, possibly anticipated at the next time step $t+\delta t$. 
This was argued to give an inadequate reflection of the cognitive heuristics employed by humans for collision avoidance \cite{moussaid2011simple}.
Karamouzas et al. \cite{Karamouzas2014universal} demonstrated that positional variables 
had better be substituted by an anticipated time to collision (TTC), which is the earliest time $\tau_{ij}\in (0,\infty)$ at which agents $i$ and $j$ are expected to collide;
 in the light of empirical data on inter-pedestrian spacings, they established an interaction potential 

\begin{equation}
    V^{TTC}(\tau)= K^{TTC} \frac{ \exp(-\tau/\tau_c)}{\tau^p},
    \label{eq:E_TTC_Karamouzas}
\end{equation}
where $p=2$ and $\tau_c$ was set to $3.0\,\mathrm{s}$.
In our case, unlike Karamouzas et al. \cite{Karamouzas2014universal,karamouzas2017implicit},  each agent $i$ will only consider the neighbor (say, $j$) in their f.o.v. with the shortest TTC and will also eschew
encroachments into their personal space (of relative width $\epsilon_i^{\star}$). 
Under some approximations (see \emph{Methods}), this leads to

\begin{equation}
    e_i^{TTC} =  \frac{1}{\epsilon_i^{\star}} \int_0^{\epsilon_i^{\star}} V^{TTC}[\tau_{ij}(\epsilon)]\, d\epsilon
    \label{eq:Ej_TTC_avg}
\end{equation}
The focus on only the shortest TTC
\blue{is consistent with findings from virtual reality experiments, where it was observed that (eye-tracked) participants tend to fixate one of the agents in the virtual crowd with the highest risks of collision
before initiating an avoidance maneuver around this person \cite{meerhoff2018guided}; this suggests that
the collision-avoidance interaction was dominated by this very person. 
More anecdotally, this focus on the most imminent collision also}
echoes Primo Levi's impression (in a wholly unrelated context \cite{levi2014if}) that one experiences fears and pains one at a time, the most acute coming first, as though the smaller ones remained hidden behind it while it persists.

TTC-based anticipation is also operational with respect to walls, but their linear shape 
modifies the technical calculation of the TTC. More specifically, the TTC $\tau_{iw}$ is defined as the shortest time after which the disk representing the agent would collide with a linear segment of a wall if it moves at the test velocity $\boldsymbol{u}$; no personal space is considered in this case because the proximity to walls is already penalized in the floor field (Sec.~\ref{sub:floor_field}).

This concludes the summary of the pseudo-energies entering the perceived cost function
\begin{eqnarray*}
\E(\boldsymbol{u}) &= &
\E^T\Big[\boldsymbol{r}(t)+\delta t \, \boldsymbol{u}\Big] +
\E^{\mathrm{personal}}\Big[\boldsymbol{r}(t)+\delta t \, \boldsymbol{u}\Big] + \\
&&
\delta t\, \Big[
e^{\mathrm{speed}}(u) + e^{\mathrm{inertia}}(\boldsymbol{u}) +
e^{\mathrm{TTC}}(\boldsymbol{u})
\Big],
\end{eqnarray*}
whose minimization, performed using a Nelder-Mead algorithm \cite{olsson1975nelder}, yields the desired velocity $\boldsymbol{u}^{\star}$, for each agent (see \fig{fig:Sketch}).

\begin{center}
\begin{table}[h]
\centering
\begin{tabular}{|c|C{4cm}|c|}
\hline 
Symbol & Definition & Value\tabularnewline
\hline 
\multicolumn{3}{c}{\bf Decision-making layer}\tabularnewline
\hline 
$\delta t$ & Decision-making time & 0.1 s\tabularnewline
\hline 
$u^{\infty}$ & Preferential speed, or free-walking speed & $ \mathcal{N}(1.4\ms,0.2)$ \tabularnewline
\hline 
$\mu$ & `Inertial' coefficient & 0.01
\tabularnewline
\hline 
$\eta$ & Repulsive coefficient associated with personal space & 0.8
\tabularnewline
\hline
$\epsilon$ & Spatial extent of the personal space (relative to body width)  & 0.2\tabularnewline
\hline
$d_c$ & Characteristic repulsion length of walls & 20~cm\tabularnewline
\hline
$\theta$ & Visual cone (half-angle)  & $70^{\mathrm{\circ}}$ \tabularnewline
\hline
\multicolumn{3}{c}{\bf Mechanical layer}\tabularnewline
\hline 
$\tau^{\mathrm{mech}}$ & Relaxation time & 0.2 s\tabularnewline
\hline 
$\nicefrac{\kappa}{m}$ & Renormalized body stiffness & $10^6$\tabularnewline
\hline 
\end{tabular}
\caption{Definitions and values of the model parameters. Note that $\delta t$ is chosen
somewhat shorter than the reaction time to complex stimuli \cite{gerson2005cortical}, because the human
decision-making process is more sophisticated than that modeled here.}
\label{tab:parameters}
\end{table}
\end{center}

\subsection{Mechanical contacts}
Collisions between pedestrians are rare but may occur in very dense crowds. Within ANDA, this will happen if the desired velocity $\boldsymbol{u}_i^{\star}$ leads to a collision within the decisional update time $\delta t$; the ensuing collisions are handled by a mechanical `layer' (\fig{fig:Sketch}), which solves Newton's equations

\begin{equation}
    \ddot{\boldsymbol{r}}_i = \frac{\boldsymbol{u}_i^{\star}-\dot{\boldsymbol{r}_i}}{\tau^{\mathrm{mech}}} 
        + \frac{1}{m} \sum_j \boldsymbol{F}_{j \to i}^{\mathrm{Hertz}}
        + \frac{1}{m} \sum_{w \in \mathrm{walls}} \boldsymbol{F}_{w \to i}^{\mathrm{Hertz}},
    \label{eq:Newton_mech}
\end{equation}
wherein agents self-propel in the direction of their desired velocity $\boldsymbol{u}_i^{\star}$, which they reach in a characteristic time $\tau^{\mathrm{mech}}\approx 0.2\,\mathrm{s}$ in free space. 
\blue{
(Note the clear connection with active matter: if $\boldsymbol{u}_i^{\star}$ were governed by an Orstein-Uhlenbeck process and fluctuated, Eq.~\ref{eq:Newton_mech} would describe interacting active Orstein-Uhlenbeck particles in the underdamped regime
\cite{caprini2021inertial}).
}
Besides, they are assumed to interact with each other via frictionless Hertzian contact forces 
$\boldsymbol{F}_{j \to i}^{\mathrm{Hertz}}=\kappa \, P\Big(\frac{\sigma_i + \sigma_j}{r_{ij}} - 1\Big) \, (\boldsymbol{r}_i - \boldsymbol{r}_j)$,
where $P(x)=\max(0,x)$, as though they were homogeneous elastic cylinders of radii $\sigma_i$ and $\sigma_j$ with parallel axes. Similarly, contacts with a wall $w$ result in a force  
$\boldsymbol{F}_{w \to i}^{\mathrm{Hertz}}= \kappa \, P\Big(\frac{\sigma_i}{r_{iw}} - 1\Big) \, (\boldsymbol{r}_i - \boldsymbol{r}_w)$, where $\boldsymbol{r}_w$ is the wall point closest to $i$ and $r_{iw}$ is the distance to the wall. Numerically, \eq{eq:Newton_mech} is solved with a velocity Verlet algorithm, using a typical time step $dt=2\cdot 10^{-4}$.

Of course, the present mechanical description could be refined in the future, turning to more realistic shapes and interactions for the agents. These improvements can easily be integrated within the sturdy theoretical ground
outlined here, which already convincingly describes many features of pedestrian dynamics, as exposed in the next section.

Interestingly, the structure of the differential equations thus obtained, namely, the combination of \eq{eq:Newton_mech} with the minimization performed in the decisional layer, 
\rev{contains some (more or less subtle) differences with those commonly used \cite{helbing1995social,moussaid2011simple,van2020generalized}, associated with the delineation of two distinct relaxation processes.}
{differs, in more or less subtle ways, from those in common use \cite{helbing1995social,moussaid2011simple,van2020generalized}; in particular, it delineates two distinct relaxation processes.  }
 These differences are not always anecdotal (\red{Appendix~D}).

\section{Results and Discussion}
\label{sec:results}

This section evinces that, despite its simplicity, the ANDA model succeeds in quantitatively reproducing empirically observed features. Most importantly, although these situations cover a wide range of contexts and  densities, no adjustment of the main parameter set (apart from very marginal ones required by the context) is needed. We will initially validate the algorithm at the individual level and then move to collective properties.

In the following simulations, we used the main model parameters detailed in Table.~\ref{tab:parameters}. Regarding the pedestrian shapes, whenever \emph{crowds} are simulated, their body radii will be chosen in a normal distribution of mean 22.5~cm
and standard deviation 2~cm. The preferential speeds will typically be normally distributed around $u^{\mathrm{\infty}}=1.4\pm 0.2\ms$, but bounded below by $1.0\ms$. 

\subsection{Collision avoidance by a single pedestrian}
\label{sub:binary_coll_avoidance}
\begin{figure}
\centering
\includegraphics[width=0.75\columnwidth]{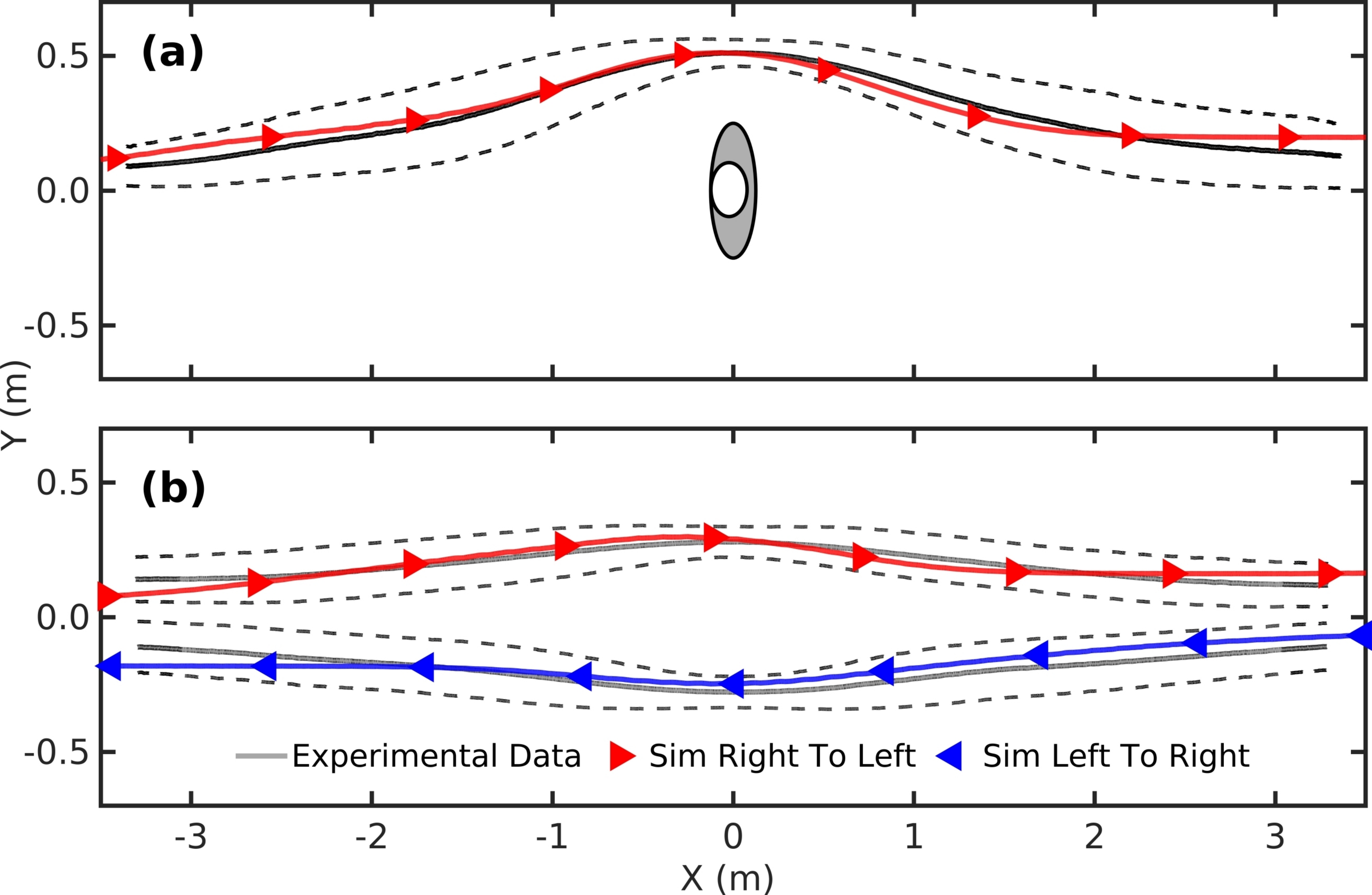}
\caption{Avoidance maneuvers in a corridor: {\bf(a)} Avoidance of a still-standing pedestrian `obstacle' at the center; {\bf(b)} trajectories of two counter-walking pedestrians. The computer simulations (coloured lines) are compared to the experiments (grey lines); the dashed lines materialize an envelope of width the standard deviation on each side of the moving average (solid line). The experimental trajectories were shifted along $Y$ to start at $Y=0$. }
\label{fig:Avoidance}
\end{figure}

To begin with, we probe binary avoidance maneuvers involving two pedestrians (body radius: $r_p=25\,\mathrm{cm}$, preferential speed: $1.4\ms$) in a 10 m length and 3 m width corridor, in two simple setups mimicking the experiments of Moussaid et al. \cite{Moussaid2009experimental}. In the first setup (Fig. \ref{fig:Avoidance}a), one pedestrian stands still in the center of the corridor, at $(X,Y)(0,0)$. Meanwhile, the other one 
is asked to cross the corridor from an initial position at $X=-5\,\mathrm{m},\,Y\in[-r_p,r_p]$ to a distant target zone centered at $(X=5\,\mathrm{m},\,Y=0 \,\mathrm{m})$, avoiding this `obstacle' \href{https://drive.google.com/file/d/1L2fGup_izpplfIDFQh0NeULRVmUrm2uK/view?usp=sharing}{(Movie S1)}. Mostly by adjusting $\mu$, we manage to obtain a close-to-perfect agreement between the model output and the average experimental behavior, apart from the avoidance-side preference (which is overlooked in ANDA and washed out of the experiments by plotting the absolute transverse displacements $|Y|$ instead of $Y$). The quality of the fit may even surpass that obtained with a social force model specifically calibrated for these experiments in the seminal original paper \cite{Moussaid2009experimental}.

In the second setup (Fig. \ref{fig:Avoidance}b), the pedestrians are initially at opposite ends of the corridor ($X=\pm5\,\mathrm{m},\,Y\in[-r_p,r_p]$) and walk in opposite directions. They start deviating from the central line already 3~meters ahead of the point of encounter (so typically 6~meters away from their counterpart, \href{https://drive.google.com/file/d/1L2fGup_izpplfIDFQh0NeULRVmUrm2uK/view?usp=sharing}{Movie S1}). The maximal transverse displacement, at the point of encounter, appears to be smaller than in the first setup with the static obstacle ($Y\approx 50\,\mathrm{cm}$), but recall that, in this second setting, the workload associated with the avoidance maneuver is shared by two pedestrians; in fact, the deviation is undertaken earlier in the second setting, in line with the expectations based on the TTC.
The (statically averaged) experimental trajectories are very well reproduced, without further adjustment compared to the first case.
\rev{}{Furthermore, we checked that the longitudinal and transverse speeds during the simulated avoidance maneuver are also consistent with the experimental ones, whereas a model featuring strong and long-ranged repulsions based on distance would predict sizable longitudinal braking.}

\rev{.}{
Going beyond controlled experimental settings, we now turn to the extensive empirical observations of binary collision avoidance passively garnered by  Corbetta et al. \cite{corbetta2018physics} in a train station.
The researchers delved into the relationship between the lateral distance before the onset of an avoidance maneuver and the lateral distance when pedestrians pass each other side by side. 
We implemented a similar scenario in ANDA with agents entering a rectangular space from opposite sides and aiming for a target on the opposite side, with some random lateral offset with respect to their entrance. The numerical results are visually quite close to the empirical ones, as depicted in \textcolor{blush}{Fig.~S6}.
}

\subsection{Many-body collision avoidances: Antipodal experiment on a circle}

Having validated the model for two-body interactions, we move on to the interactions among a larger number of people. Circle antipodal experiments are an archetypal way to probe many-body collision avoidances. In this configuration, pedestrians are initially positioned on a circle (of 5 or 10~m of radius for example \cite{xiao2019investigation}) with uniform spacing between them and instructed to quickly reach the antipodal positions as soon as a signal is heard. 
Were they to walk straight to their targets at equal speeds, they would meet at the centre of the circle. Instead, experiments have shown that they deviate from the plain straight path soon after setting in motion. By adopting various collision avoidance and detour strategies, they manage to reach their target without substantial near collisions \cite{xiao2019investigation}. 
We replicated the experimental setup numerically for a set of 10 agents with equal preferential speeds. Without mutual interactions, these agents would walk straight ahead at very similar speeds,
despite their different directions, thanks to our use of a hexagonal lattice for the floor field (as we explained in Sec.~\ref{sub:floor_field}). Instead, we observe that they deviate from the straight path and easily manage to reach the antipodal position \href{https://drive.google.com/file/d/1XZL9ozB49iPGLCvTdyQIYNUbDNEucS3w/view?usp=sharing}{(Movie S2)}. 
The simulated trajectories are smooth and, as in the experiments, they do not excessively concentrate at the conflict-rife center of the circle; this is at odds with the output of the social force model, either in its traditional version or in a specifically designed variant, which were both deemed to significantly differ from the experimental results in \cite{xiao2019investigation}. 

\begin{figure}
\centering
\includegraphics[width=0.5\columnwidth]{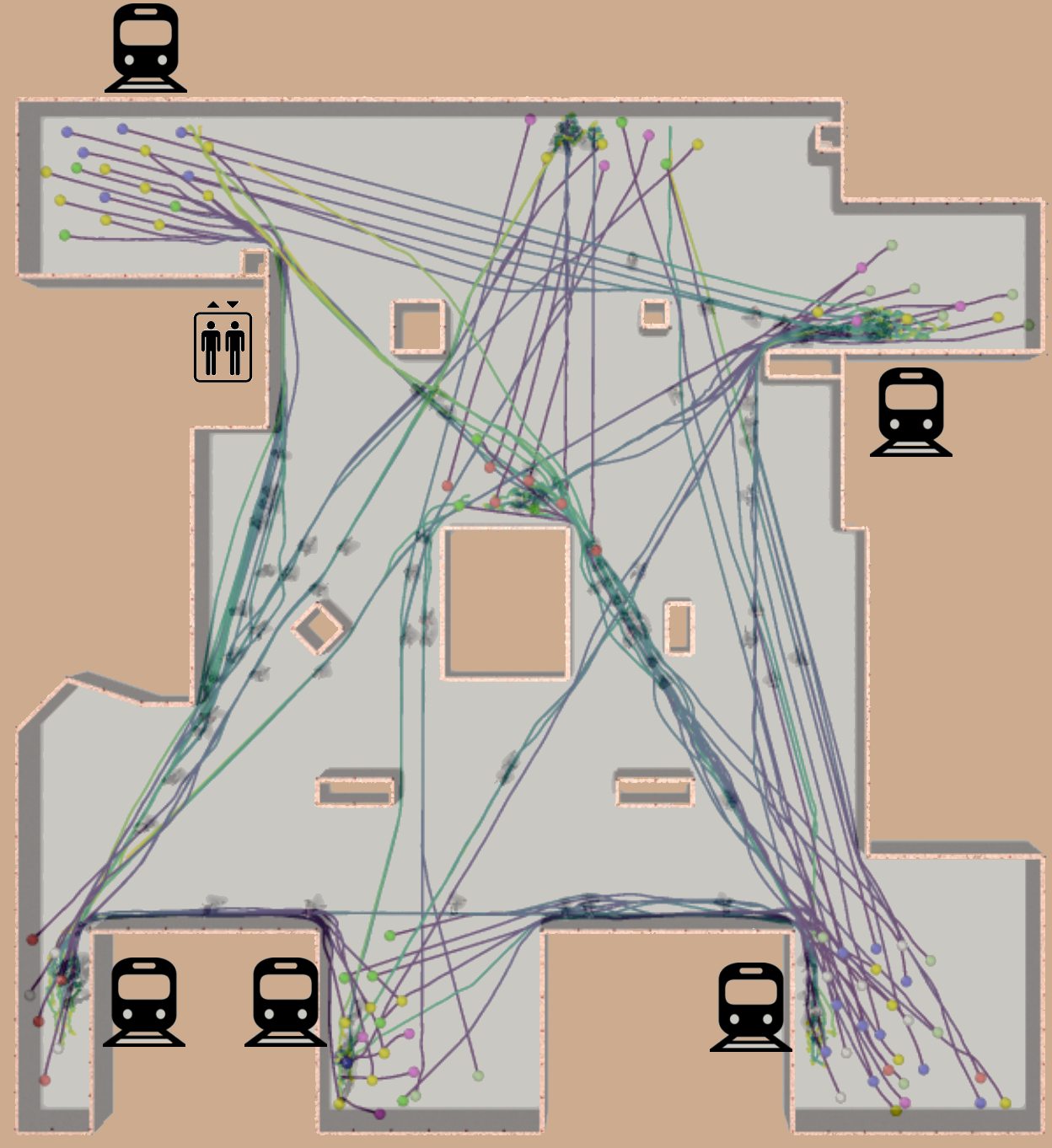}
\caption{Navigation of $\sim 100$ agents, split into 7 groups with distinct target zones,  in a complex geometric layout inspired by the ground floor of the Montparnasse train station in Paris, France. The initial positions are marked with colored disks and the solid lines that represent the trajectories evolve from blue to yellow with time.}
\label{fig:Montparnasse}
\end{figure}

\subsection{Navigation in a complex geometry}
\label{sub:complex_geometry}

Navigation in a complex environment adds another layer of complexity to the foregoing multi-agent
scenarios, in that each agent must also interact with the built environment. To characterize pedestrian flows in these (practically relevant) situations, we designed a geometry
inspired by the ground floor of Montparnasse train station in Paris, France, in which about 100 pedestrians (walking at $u^{\infty}\geqslant 1.2\ms$),
classified in 7 groups with distinct targets, were simulated. The simulation runs in a matter of minutes on a single CPU core. Thanks
to the use of a floor
field to attract each group to their specific target, the agents make sensible route
choices to their destination (Fig.\ref{fig:Montparnasse}), without any need for an additional tactical layer. In particular, they go around walls and obstacles whenever needed and are not constrained to strictly adhere to a predefined path.
Furthermore, the simulated dynamics (shown in \href{https://drive.google.com/file/d/1fy2morxISAN6OxCAYTgxBmD8eIgvy0dg/view?usp=sharing}{Movie S3}) are qualitatively convincing at the local level, as far as one can judge with the naked eye: 
the agents succeed in navigating toward their targets in a realistic way, generally avoiding collisions with their counterparts and the walls. 
Still, even closer inspection of the video reveals some hesitancy in the central region when an agent endeavors to cross a group of static pedestrians who have reached their target. This particular situation is further discussed in \red{Appendix~E}.


\subsection{Unidirectional flow}
\label{Uni_flow}

Turning to higher densities, we investigate the effect of density on unidirectional flow by means of the \emph{speed-density} relation, a broadly used quantitative benchmark for models of pedestrian dynamics. A corridor of length $L_X= 16\,\mathrm{m}$ and width $L_Y= 3\,\mathrm{m}$ is considered (similar to the experimental scenario in \cite{zhang2011transitions}), with periodic boundary conditions (PBC) in the horizontal direction. The number of pedestrians inside the corridor is varied from 12 to 144, thus achieving densities ranging from 0.25 to 3~ped/m$^2$. Each simulation runs for 100 seconds, the last 75~s of which are
used to compute the speed-density relation by averaging the speeds of all agents.

\begin{figure*}[htp]
\includegraphics[width=\textwidth]{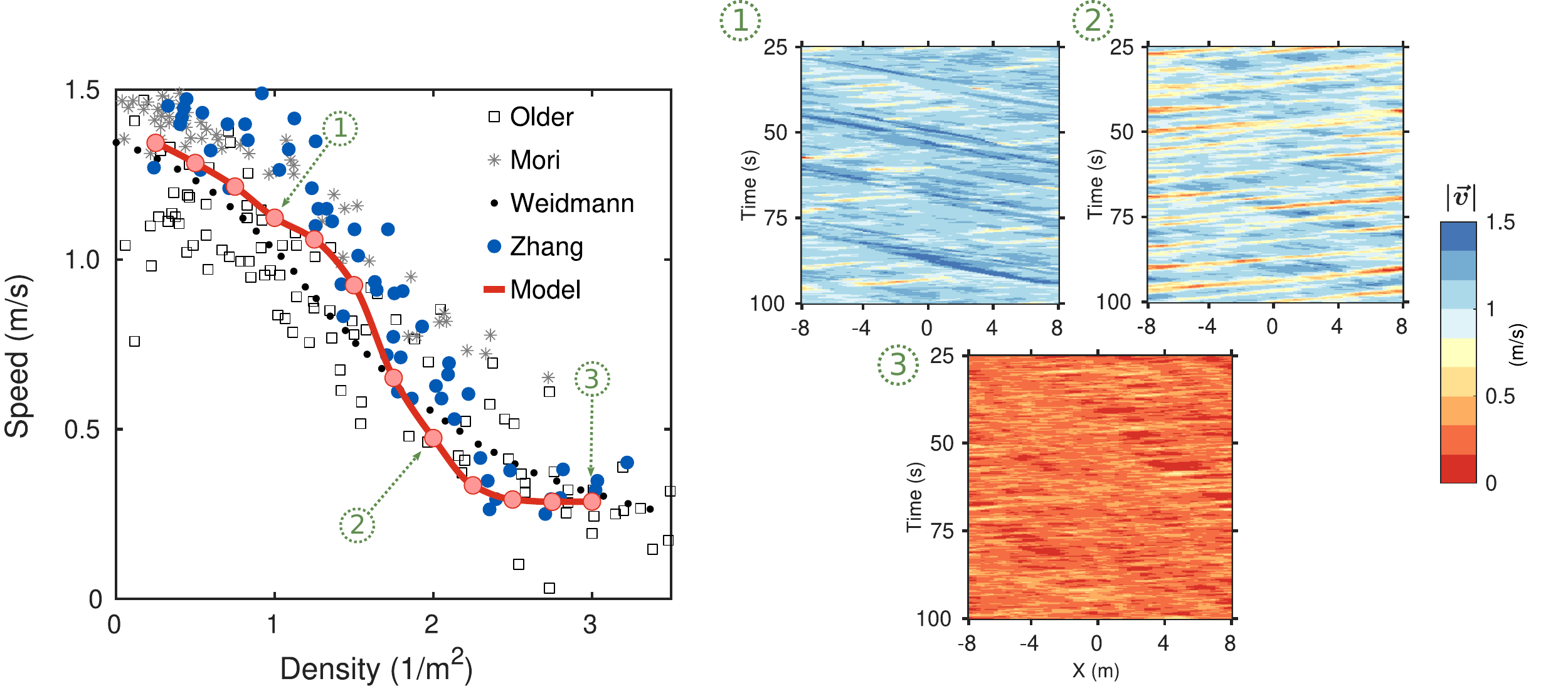}
\caption{Unidirectional flow along a corridor of (periodic) length $L_X=16\,\mathrm{m}$ and width $L_Y=3\,\mathrm{m}$. (Left) Variation of the simulated mean pedestrian speed with the density, shown along with various experimental data sets (Older \cite{Older1968}, Mori and Tsukaguchi \cite{Mori}, Weidmann \cite{Weidmann1993} and Zhang et al. \cite{zhang2011transitions}). (Right) Spatio-temporal diagrams of the coarse-grained local speed, represented at different densities, as labeled in the left panel.}
\label{fig:Speed-density}
\end{figure*}

The numerical outcome in \fig{fig:Speed-density} follows the same trends as the empirical data, with a monotonic decay of speed with density that gets sharper around 1.5-2~ped/m$^2$. 
The situation at high density $\rho>2.5\,\mathrm{ped}/m^2$ deserves additional comments. First, the speed is still nonvanishing in this regime, consistently with controlled laboratory experiments \cite{jin2019observational} as well as empirical observations even at (much) higher densities, (far) above 6~ped/m$^2$ in the pilgrim processions during the Hajj \cite{helbing2007dynamics}. Secondly, in the model, the speed seems to level off, possibly excessively. In this regime, pedestrian shapes start to become overly important and,
not so surprisingly, the current approximation of pedestrians as disks (to dispose of rotational degrees of freedom) then reaches its limits. 

The spatio-temporal diagrams of speed shown in the lateral panels of \fig{fig:Speed-density} for different densities shed light on finer details about the flow dynamics, notably the occurrence of stop-and-go waves in the higher-density regime. 
This type of instability, wherein a slow or `jammed', dense phase emerges locally in the unidirectional crowd and propagates upstream, is routinely observed in various forms of traffic and in pedestrian motion \cite{portz2011analyzing} when the average spacing between agents is reduced. In our simulations, no stop-and-go wave was observed at the lowest considered density (label (1), $\rho=1\,\mathrm{ped/m^2}$), in line with the experiments; instead, clusters of fast-walking agents create dark blue streaks, that move downstream at their (high) walking speed (\href{https://drive.google.com/file/d/1QiTdEIZapVgtxrpLH9ULeKshqqyc6_CA/view?usp=sharing}{Movie S4}). When the density in the corridor increases (label (2), $\rho=2\,\mathrm{ped/m^2}$), the spatio-temporal diagrams are qualitatively altered, with regions of halted pedestrians propagating upstream, corresponding to stop-and-go motion (\href{https://drive.google.com/file/d/1QiTdEIZapVgtxrpLH9ULeKshqqyc6_CA/view?usp=sharing}{Movie S4}).

Interestingly, these backward-travelling waves are conspicuous in our coarse-grained diagrams of speed, but are not apparent when we plot the coarse-grained diagrams of density \red{(Fig.~S3)}.
Thus, the density indicator used to experimentally detect stop-and-go motion in \cite{jin2021pedestrian} appears to be less telling than the (longitudinal) speed, as we elaborate in \red{Appendix F}. Finally, at even higher densities (label (3), $\rho=3\,\mathrm{ped/m^2}$), promiscuity slows the flow even more and the dynamics become globally more hampered, but also less bursty.

\begin{figure}[h]
\centering
\includegraphics[width=0.6\columnwidth]{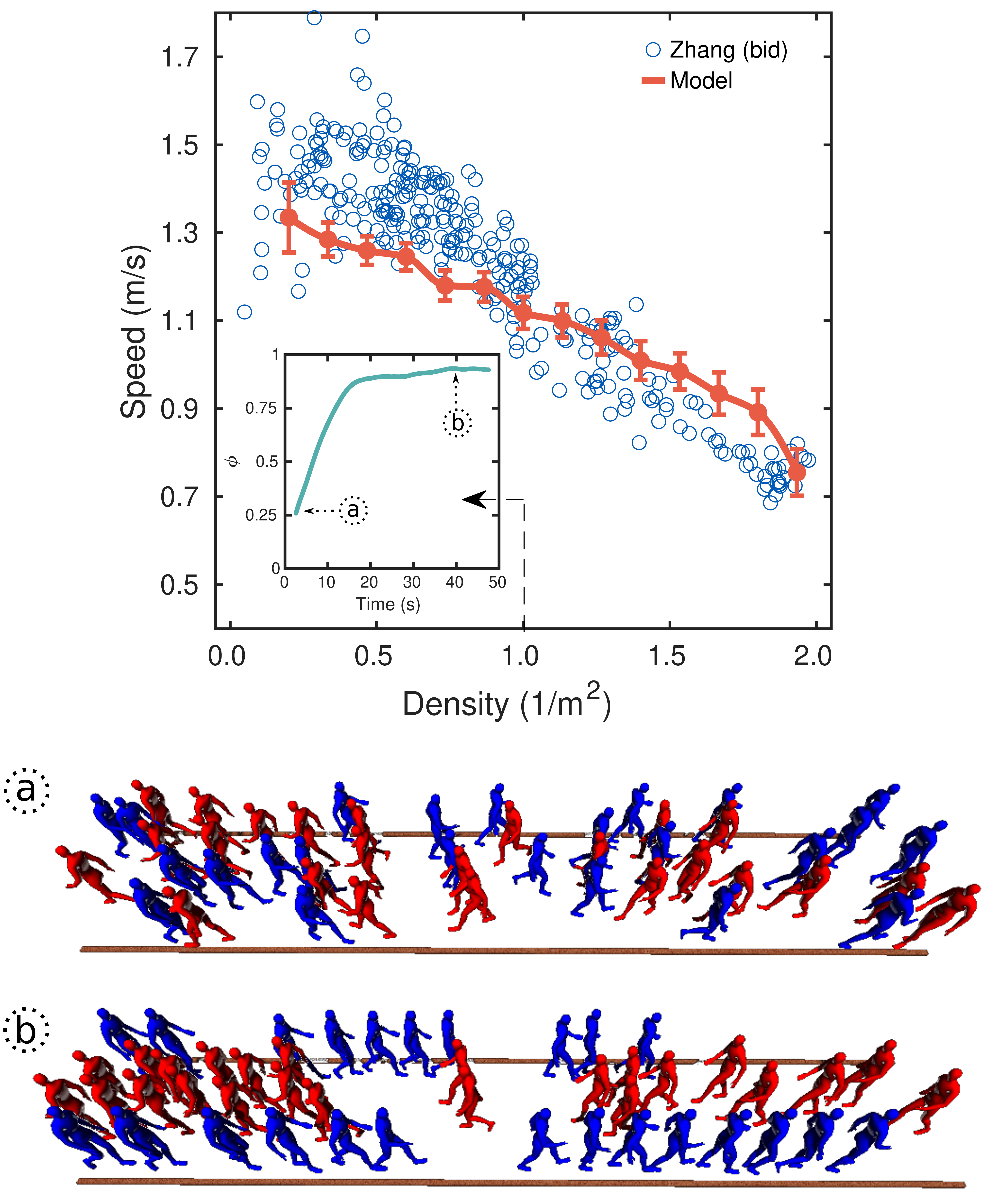}
\caption{Variation of the average pedestrian speed with the mean density in a bidirectional flow corridor. (\emph{Inset}) Time evolution of the order parameter $\phi$ (defined in \eq{Phi_param}) for the characterization of Lane-formation phenomena. The letters (a) and (b) indicate the times corresponding to the images below. (Bottom) Snapshots of the crowd at different times, rendered with the CHAOS software developed by INRIA: (a) well-mixed crowd at the beginning of the simulation, (b) crowd structured in counter-walking lanes. Refer to the main text for the dimensions of the corridor.}
\label{Lane_Formation}
\end{figure}




\subsection{Bidirectional flow}
\label{sub:bidirectional}

Bidirectional flows are also ubiquitous in daily-life and exhibit particular features.
Regardless of the type of facility, the system will most probably evolve into a segregated state where people end up forming lanes \cite{navin1969pedestrian,yamori1998going}. Such organization reduces collision risks with counter-walking agents and allows people to walk faster. While lane formation has historically been a major benchmark test for any new model, it is noteworthy that this phenomenon is not specific to pedestrian crowds, but is widely found in other active assemblies, and even in simply driven particle systems, such as colloids \cite{dzubiella2002lane}. Indeed, it is underlain by a generic mechanism: in the non-organized flow, agents undergo small transverse moves after each collision; upon aligning behind someone walking in the same direction, these collisions become much less frequent and the aligned state is thus stabilized.
Still, the lane-forming state cannot be reached within some pedestrian models in crowded corridors. This deficiency, leading to deadlocks, was underlined in \cite{xu2021anticipation} and remedied by an \emph{ad hoc} anticipation mechanism. Here, we show that such additions are not required by ANDA and that it can natively describe bidirectional flows.

Filling the previously defined corridor with an equal number of left and right-moving agents, we find
a \emph{speed-density} relation comparable with that obtained for unidirectional flow and consistent with experimental measurements, as shown in the main panel of \fig{Lane_Formation}. The process of lane formation is unveiled by plotting the evolution of a lane-order parameter $\phi \in [0,1]$ (see \emph{Methods}) with time in the inset of \fig{Lane_Formation}, for $\rho=1\,\mathrm{ped/m^2}$. 
We observe that it gradually increases from $\phi \simeq 0$ in the early stages [label (a)], denoting a disordered state, to $\phi \simeq 1$ [label (b)], indicating a laned structure where pedestrians are segregated, and the transition takes about 10 to 15 seconds. This simulated lane-formation time for $\rho=1\,\mathrm{ped/m^2}$ lies just in-between the values measured experimentally for $\rho=1\,\mathrm{ped/m^2}$ and $\rho=2\,\mathrm{ped/m^2}$ \cite{jin2019observational}.

\subsection{Bottleneck flow and evacuations}
\label{sub:evacuation}

Introducing a narrowing (a bottleneck) across the corridor results in converging streamlines, hence possible clogs if the flow is dense and 
the bottleneck is not much wider than a few `particles'. Clogging may occur when the particles are grains or animals \cite{Zuriguel2014clogging}, but is vested with special interest
for pedestrians, for it may be critical during egresses or evacuations under emergency conditions. The topic has thus received much attention and some paradoxical effects have been brought to light: While more haste often makes the evacuation quicker, in very competitive settings, higher individual preferential speeds may be counterproductive, leading to long-lived clogs, observed empirically as well as experimentally. This is the well-known `faster-is-slower' effect (FIS), first predicted numerically \cite{helbing2000simulating} and then demonstrated experimentally in a variety of assemblies \cite{Pastor2015experimental,Zuriguel2014clogging}.

\begin{figure*}[h]
\includegraphics[width=1.0\textwidth]{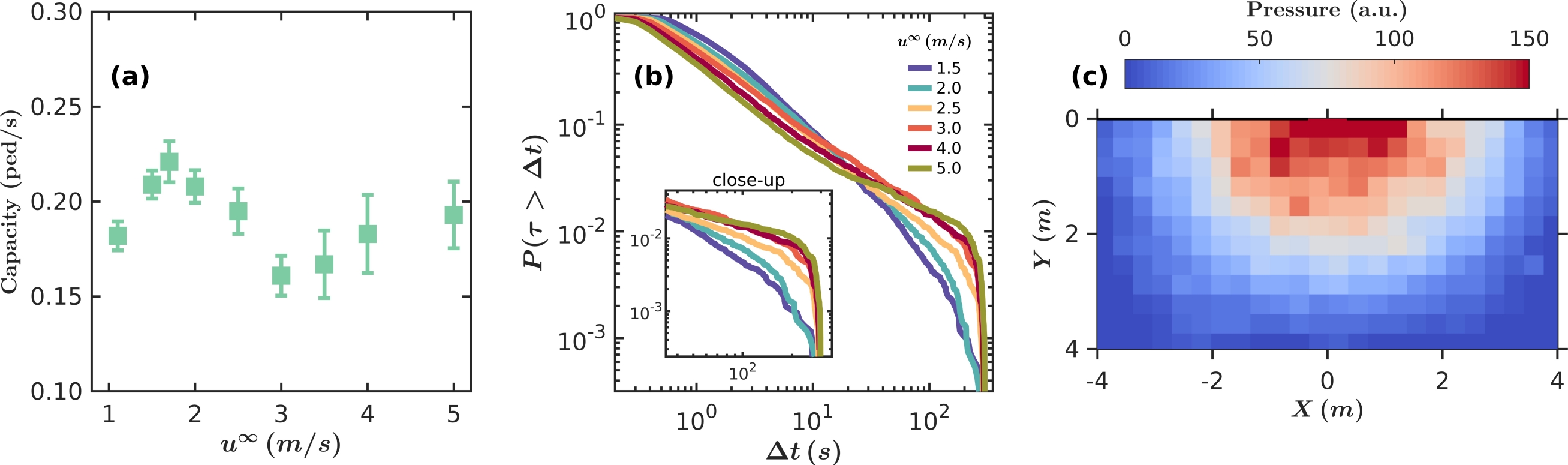}
\caption{Evacuation dynamics through a narrow doorway of width $w=60\,\mathrm{cm}$, centered at $(0,0)$. {\bf(a)} Dependence of the exit capacity on the preferential speed $u^{\infty}$. {\bf (b)} Survival function $P(\tau>\Delta t)$ of time gaps between successive egresses. (\emph{Inset}) Detailed view of the main panel to identify the relative significance of the distributions' tails. {\bf (c)} Average pressure field due to physical contacts between agents in the vicinity of the exit, for $u^{\infty} = 3\ms$.} 
\label{fig:evac}
\end{figure*}

Here, we simulate an evacuation from a rectangular room initially filled with 150 randomly positioned agents who strive to egress through a doorway of width $w$ in the middle of one wall. The agents' eagerness to evacuate affects the terminal cost $\E^{T}$ and, as a consequence,
their preferential speed $u^{\infty}$; the latter will be varied and used as a proxy for eagerness. To avoid deadlocks at the bottleneck, every second the preferential speed of each agent
undergoes a small random fluctuation (drawn from a normal distribution of standard deviation 0.2) around its initial, agent-dependent value.
Each simulation is replicated around 100 times to collect sufficient statistics to overcome the expected strong fluctuations; for scientific rigor, but with virtually no impact on our results,  the first and last egresses in each realization were discarded, to focus on the quasi-stationary state.

Gauging the evacuation efficiency by the exit capacity $J_s$, i.e., the pedestrian throughput, we show in \fig{fig:evac}a that the FIS is retrieved when the door is very narrow, $w=60\,\mathrm{cm}$: the capacity plummets as soon as $u^{\infty}$ exceeds $1.7\ms$. In this competitive regime,
the total evacuation time $T_{\mathrm{evac}}$ exhibits strong fluctuations, rationalized by the (infrequent, but not sporadic) occurrence of very long clogs. These clogs generate large time gaps $\tau_i$ between successive egresses,
which impact $T_{\mathrm{evac}}= \sum_i \tau_i$. The survival functions $P(\tau>T)$ of the $\tau_i$, represented in a logarithmic plot in \fig{fig:evac}b, are suggestive of a power-law-like behavior with heavy tails, whose slopes get flatter and flatter (\emph{Inset}) as agents get more and more hurried ($u^{\infty}>2-3\ms$), in contrast with the fast
decays observed for more placid agents and/or wider doors \red{(Fig.~S4f)}. These features are
in perfect agreement with previous experimental findings \cite{Pastor2015experimental}.  Beyond $u^{\infty}\simeq 3\ms$, the flattening trend gets less visible, as the stronger pushing forces counter the clogging phenomenon \cite{sticco2017beyond}. In parallel, the \emph{typical} time gaps is reduced as the agents move faster; therefore,
the capacity starts increasing again with $u^{\infty}$ (\fig{fig:evac}a).

For a slightly larger door, of width $w=70\,\mathrm{cm}$, the FIS is still noticeable, but not as conspicuous: the nonmonotonicity of $J_s$ is only tentatively seen around $u^{\infty}=4\ms$ \red{(Fig.~S4a)}.
This is broadly in line with the findings of \cite{Pastor2015experimental}, where the FIS was clearly present, but fairly small for a 69 cm-wide door. 
For wider doors, $w>70\,\mathrm{cm}$, 
the FIS fades away from our simulations \red{(Fig.~S4b,c)}: Higher preferential speeds $u^{\infty}$ lead to higher throughputs.

More quantitatively, the variations of the \emph{specific} capacity $J_s$ with the door width closely matches experimental reports, down to narrow doors \red{(Appendix~G)}. For \emph{very} narrow doors,  the model reaches its limits. \blue{This is primarily due to the description of agents as frictionless disks, which overlooks important factors such as shearing stress, cross section or body orientation. These features actually play a fundamental role when it comes to correctly characterizing how a person moves through a narrow space. 
Therefore, the prospective refinement of the mechanical layer will probably further enhance the realism of the description of flows through narrow bottlenecks.} 


Finally, thanks to the neatly delineated mechanical layer, contact forces can be defined rigorously, by contrast with models mingling pseudo-forces and real forces \rev{}{whenever the latter are not calibrated independently}. More precisely, the contact force $F_{j \leftarrow i}$ exerted by $i$ on $j$ reads $F_{j \leftarrow i}=-\frac{d U_2}{\boldsymbol{r_{ij}}}$, where $U_2$ is the Hertzian potential defined previously. If one overlooks variations in the agent's surface area $\mathcal{A}$, the pressure can then be defined, from a continuum mechanics standpoint, as the sum of contact forces exerted on $j$ divided by $\mathcal{A}$, viz.,  $\sum_i F_{ij} / \mathcal{A}$. With this definition, \fig{fig:evac}c shows the average pressure field during the evacuation, i.e., the mean pressure felt by agents at each position in space. It is noteworthy that these pressure fields look similar to the density fields measured experimentally in evacuations under similar competitiveness \cite{zuriguel2020contact}.

\subsection{Effect of distracted pedestrians}
\label{sub:distracted}

So far we have shown that key pedestrian dynamics features could be replicated in various settings with a single set of model parameters. Now, we purport to show that the sound physical basis of ANDA enables us to extend it to an even wider range of situations by straightforwardly adapting its parameters. 

To illustrate this, we consider the effect of digital distraction \cite{nasar2013pedestrian,ropaka2020investigation,larue2022prevalence}. In our increasingly connected societies, with the advent of the Internet of things, more and more pedestrians are indeed looking at their smartphones (or other connected devices) while walking; even near road crossings, more than one pedestrian out of six may be involved in a such an activity (17\% in a 2020 study in Athens, Greece \cite{ropaka2020investigation}). The ensuing distraction impacts their navigation in that it impairs their situational awareness \cite{nasar2008mobile,nasar2013pedestrian}, especially when texting or web-browsing \cite{ropaka2020investigation,jiang2018effects}. Their walking speed is then reduced \cite{ropaka2020investigation,jiang2018effects}, as is their eye scanning frequency (by upwards of 25\% in controlled outdoor experiments with college students \cite{jiang2018effects}). 
The whole topic has gained serious practical relevance as `smartphone-walking' has entailed a sharp rise in pedestrian injuries. Already in 2010, of the thousands of pedestrians killed in traffic accidents in the US, 3.7\% were engaged in a mobile phone activity, as compared to 0.6\% in 2004 \cite{nasar2013pedestrian}; the numbers have most probably considerably risen since then, further heightening societal concerns, notably in Japan \cite{NHK2021japan}. 

\begin{figure*}
\includegraphics[width=1\textwidth]{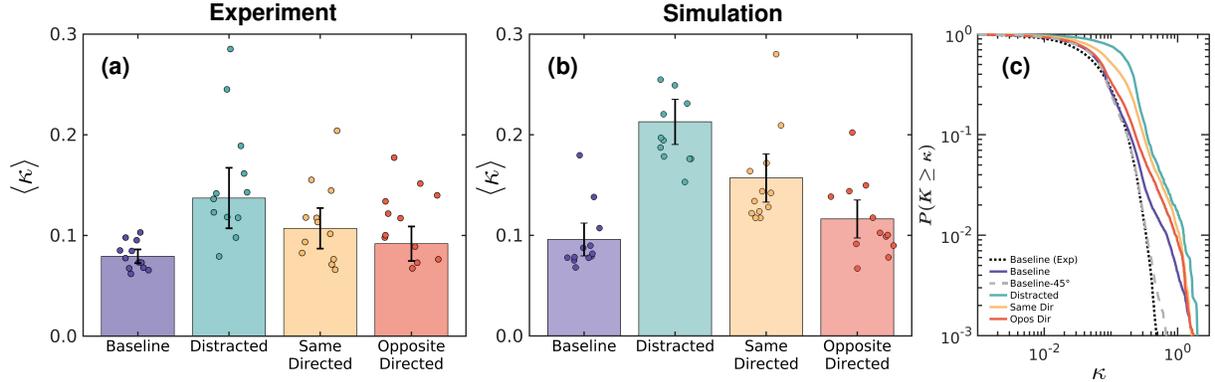}
\caption{Impact of distracted pedestrians on the chaoticity of the flow: {\bf(a)} Suddenness of turn $\kappa$ in the experiments of \cite{murakami2021mutual}, distinguished between a reference case without distracted agents (`baseline'), the distracted agents, the non-distracted people walking in the same direction and those walking in the opposite direction; {\bf(b)} Suddenness of turn $\kappa$ computed in our model. {\bf (c)} Survival functions $P(K>\kappa)$ in the reference experimental case and for different groups of agents in the simulations. The dashed grey line refers to a simulation in which the half-angle of the visual field was reduced to $\theta=45^{\circ}$.}
\label{fig:Murakami}
\end{figure*}

To explore the effect of digital distraction in a broad range of situations, in particular on streets, numerical models are of great avail. In order to account for it, we notice that distraction through screens, albeit a complex psychological process, mainly entails that agents less frequently refresh (update) their perception of their surroundings and adapt their motion to it \cite{nasar2013pedestrian}. Indeed, in a collision avoidance scenario, they tend to turn at the last
moment, with a delay of around 0.5 or 1 second in their response 
\blue{(i.e., very roughly, twice or thrice the normal reaction time to a complex visual stimulus \cite{gerson2005cortical})},
 compared to the reference case  \cite{murakami2022spontaneous}. This latency effect can readily be transcribed into ANDA by 
 \blue{the decisional update time interval $\delta t$ of distracted agents thrice as large as that of
 standard agents  ($\delta t=0.3\,\mathrm{s}$ instead of $0.1\,\mathrm{s}$; recall that the value of $\delta t$ chosen in the model is lower than the real one because the simulated decision-making is much coarser)};
 the walking speed of distracted agents (empirically slower than their counterparts \cite{jiang2018effects}) is set to $u^{\infty}=1.2\ms$. Of course, distraction may have secondary effects that would impact the perceived cost, but these are not addressed in this analysis.

We now test to what extent this numerical account of distraction is faithful to the experimental observations. To that end, we make use of Murakami et al.'s recent experiments on bidirectional flows in the presence of three digitally distracted agents, i.e., three participants who were instructed to use their smartphones while walking \cite{murakami2021mutual}. The researchers
observed that their presence  hampered lane formation and made the flow more chaotic, especially when the three participants 
were positioned at the front of the group of participants moving in one direction.
This is well captured by an observable $\kappa$ (detailed in the \emph{Methods}) that measures how suddenly pedestrians had to turn using the curvature of their trajectories 
\rev{}{multiplied by their instantaneous speed}: 
larger $\kappa$ values, hence a more chaotic flow, are observed experimentally in \fig{fig:Murakami}a for distracted pedestrians, but also for
the participants walking behind them in the same direction (referred to as `Same Directed')
and, to a lesser extent, those walking opposite to them (`Opposite Directed'), compared to the reference situation with no distracted agents (`Baseline').

Simulating an identical setup, as soon as the effect of head sways on $\kappa$ is accounted for by superimposing oscillations onto the simulated trajectories (see \emph{Methods}), we recover the average flow chaoticity $\kappa$ measured experimentally in the reference situation (\fig{fig:Murakami}b). 
By introducing smartphone-walking pedestrians in the front row of the crowd, the substantially enhanced chaoticity $\kappa$ observed experimentally not only for the (few) distracted people but also for the others (particularly those walking in the same direction) is surprisingly well captured by our model, wherein digital distraction mostly boils down to having a longer time $\delta t$ between updates of the desired velocity (i.e., perception of, and reaction to, the environment): 
\fig{fig:Murakami}b testifies that the trends and the variations between the pedestrians depending on their status match the experimental findings. 
To get a sharper focus, we analyze the distributions of $\kappa$ and show the survival functions $P(K>\kappa)$ in \fig{fig:Murakami}c. The exacerbated chaoticity is mostly due to the more frequent occurrence of very sudden turns (associated with large $\kappa$). Besides, while we chose to keep the very same set of model parameters throughout the main text, we noticed
that using a narrower visual field for the agents, i.e., reducing the half-angle $\theta$ from $70^{\circ}$ to $45^{\circ}$, yielded a better match between experiments and simulations (compare the purple line with the dashed grey line in \fig{fig:Murakami}c); this is also true for the mean values of $\kappa$ in the reference situation as well as in presence of distracted agents \red{(Fig.~S5)}.

\section{Conclusions}
To summarize, we have put forward a model for pedestrian dynamics that better distinguishes the psychological processes at play from the mechanical ones. In particular,
the selection of a desired velocity by each (autonomous) agent is entrusted to a decision-making layer, which 
optimizes a perceived cost, whereas physical contacts are handled with Newton's equation of motion. Many model parameters can be adjusted based on existing empirical data.
Despite the limited number of parameters left for adjustment,
the model succeeds in reproducing 
a variety of experimental features over an impressively broad range of situations and densities (without resorting to more specific adjustments, compared to other approaches), 
overcoming the need for a specific calibration in each regime.
These situations, listed in \red{Table~S1}, include collision avoidance between several agents, the \emph{speed-density} relations for unidirectional and bidirectional flows, bottleneck flows, and navigation in a complex geometry.
The model can even replicate more exotic phenomena,  which data-driven approaches would have struggled to capture, due to the lack of data. Digital distraction through smartphones, which has grown into a major issue for pedestrian safety, is one of them.

The transparent construction of the model also highlights the approximations that were made and that would need to be improved for a more faithful description of some scenarios, such as the pedestrian shape, at high densities, and the short-time approximation of the utility function, in situations where the operational dynamics include relatively far-sighted anticipation usually assigned to the tactical level. 

\blue{More broadly speaking, a model that decouples the decision-making and physical processes has quite promising implications when it comes to representing different pedestrian attributes, such as impaired motor control due to injuries or carrying baggage, or reduced decision-making ability due to various forms of distraction or conversations with others. Exploring these aspects further and testing the model's applicability in real-world scenarios looks like a promising avenue for future research.}

\section{Methods}

\subsection*{Static floor field}
The shortest-path distances $\mathcal{D}$ to the destination, entering the static floor field via \eq{eq:ET_D}, are obtained by solving
the Eikonal equation (generally used for ray tracing, \rev{}{but also in \cite{cristiani2023all}}) on a hexagonal lattice (the dual of a honeycomb lattice) before runtime and stored in memory. This is achieved by means of Dijkstra's algorithm,  considering the nearest two neighbors of each node and  evaluating  the cost $n(\boldsymbol{r})$ of traveling along an edge at the next (rather than current) node.  The use of a highly symmetric hexagonal lattice strongly curbs the spurious anisotropy that is known to be generated by the Dijkstra algorithm (e.g., on square lattices); in practice, the variations of the free walking speed with the direction of motion are reduced to less than 10\%. 

\subsection*{TTC energy accounting for encroachments into personal spaces}
Interactions based on TTC, instead of distances, are central in the model.
Let us first handle agents $j$ as hard disks of radii $\sigma_j$ and agent $i$; assuming that
all neighbors $j$ maintain their current velocities, the TTC
$\tau_{ij}$ can readily be
calculated as \cite{karamouzas2017implicit}

\begin{equation}
    \tau_{ij}(\boldsymbol{u}_i)= \frac{- \boldsymbol{x}_{ij}\cdot \boldsymbol{v}_{ij} - \sqrt{\Delta}}{v^{2}_{ij}}
    \label{eq:tau_discr}
\end{equation}
if $\Delta=(\boldsymbol{x}_{ij}\cdot \boldsymbol{v}_{ij})^2 - v_{ij}^2[x_{ij}^2 - (\sigma_i + \sigma_j)^2]\geqslant 0$, or $\infty$ otherwise. Here, $\boldsymbol{x}_{ij}$ and $\boldsymbol{v}_{ij}$ are the relative positions and desired velocities of $i$ with respect to $j$.

This equation only penalizes desired velocities that lead to physical contact. However, pedestrians are also eager  to avoid encroachments on their personal spaces. Suppose that the personal space is a disk of radius $(1+\epsilon) \sigma_i$, where $\epsilon\geqslant 0$ and $\sigma_i$ is the body radius of the agent ($i$, here). \Eq{eq:tau_discr} can then be applied to retrieve the TTC of personal spaces, $\tau_{ij}(\epsilon)$, provided that the sum of body radii $\sigma_i + \sigma_j$ is multiplied (`inflated') by  $1+\epsilon$. We recall that agent $i$ will only consider the neighbor $j$ in the f.o.v. with the most imminent TTC.
Since the transition into the personal space is actually smooth, the TTC energy should read

\begin{equation}
    e_j^{TTC}=  \frac{\int_0^{\infty} V^{\mathrm{rep}}(1+\epsilon)\,  V^{TTC}[\tau_{ij}(\epsilon)]\, d\epsilon}
    {\int_0^{\infty} V^{\mathrm{rep}}(1+\epsilon)\,   d\epsilon}.
\end{equation}

In practice, we approximated the foregoing formula by
\begin{equation}
    e_i^{TTC} =  \frac{1}{\epsilon_i^{\star}} \int_0^{\epsilon_i^{\star}} V^{TTC}[\tau_{ij}(\epsilon)]\, d\epsilon
    \label{eq:MatMet_Ej_TTC_avg}
\end{equation}
Here, $\epsilon_i^{\star}=\min\Big(\epsilon^{\star}, \epsilon_{i} \Big)$ may be lower than the maximal extent of the personal space, $\epsilon^{\star}$, if this value leads to an overlap of the personal sphere at the present time; in that case, $\epsilon_i^{\star}$ is capped to $\epsilon_i$, the largest inflation factor guaranteeing that agent $i$'s personal space does not currently overlap any other agent's. A similar averaging procedure was put forward in \cite{forootaninia2017uncertainty} to smooth the TTC energy (with respect to positions), but it was then interpreted as the result of uncertainty on the evaluation of body sizes.

If no collision with agent $j$ is ever expected, even with maximally inflated radii, i.e., $\tau_{ij}(\epsilon_i^{\star})=\infty$, then $e_i^{TTC}=0$. Otherwise, the minimal inflation leading to collision, $\epsilon_{ij}^{c} \geq 0$, can easily be derived from \eq{eq:tau_discr}. On this basis, \eq{eq:MatMet_Ej_TTC_avg} was further approximated by

\begin{equation}
    e_i^{TTC} \approx 
    \frac{\epsilon_i^{\star}-\epsilon_{ij}^{c}}{\epsilon_i^{\star}}\, V^{TTC}\Big[\tau_{ij}\Big(\frac{\epsilon_i^{\star}+\epsilon_i^{c}}{2}\Big)\Big].
\end{equation}

\subsection*{Lane formation order parameter}
To quantify the ordering in lanes in bidirectional flows, we took up the order parameter $\phi$
introduced in \cite{chakrabarti2004reentrance,dzubiella2002lane},
\begin{equation}
    \phi = \frac{1}{N}\sum_{i=1}^N \phi_i \in [0,1]\text{ with }\phi_i = \left( \frac{N_i^{Same} - N_i^{Diff}}{N_i^{Same} + N_i^{Diff}}\right)^2,
    \label{Phi_param}
\end{equation}

where $N_i^{Same}$ and $N_i^{Diff}$ are the number of pedestrians walking on the same line as pedestrian $i$, respectively in the same direction and in the opposite one, viz.,

\begin{equation}
\begin{split}
N_i^{Same} = \lbrace j, \left|y_j-y_i\right| < 3\sigma_i/2 \text{ and } \hat{v}_i \cdot \hat{v}_j > 0  \rbrace\\
N_i^{Diff} = \lbrace j, \left|y_j-y_i\right| < 3\sigma_i/2 \text{ and } \hat{v}_i \cdot \hat{v}_j < 0 \rbrace
\end{split}
\end{equation}

where we used $3\sigma_i/2$ as our characteristic length scale (remember that $\sigma_i$ is the particle radius), as supported by previous works \cite{xu2021anticipation}. Thus, $\phi$  ranges from 0 in the fully disordered (mixed) state) to 1 if the crowd is fully stratified in lanes.

\subsection*{Suddenness-of-turn parameter}
To gauge how chaotic a bidirectional flow can become with digitally distracted agents, Murakami and co-workers defined a suddenness-of-turn $\kappa$ \cite{murakami2021mutual}, which we somewhat amended here to ensure its invariance under global rotations of the frame, viz., 
\begin{equation}
    \kappa(t) \hat{=} \left\Vert \frac{\boldsymbol{e}(t+\Delta t)-\boldsymbol{e}(t)}{\Delta t} \right\Vert,
\end{equation}
where $\boldsymbol{e}(t)$ is the direction of motion at time $t$; $\kappa$ tends to the geometric curvature, 
\rev{}{multiplied by the current speed $v(t)$,}
in the limit of small $\Delta t$. For this particular work, the value of $\Delta t$ has been set to 1, although lower values were also tested to check for convergence.

$\kappa$ measures how abruptly people have to turn, and thus deviate from their preferred straight trajectory, but it is also quite sensitive to the lateral swaying motion exhibited by pedestrians' heads. This characteristic feature pertains to the mechanism of biped locomotion and it is usually neglected in simulations. However, since these gait-induced oscillations impact $\kappa$, we characterized them in Murakami's experiments \cite{murakami2021mutual} by measuring their period $\omega=1.6\,\mathrm{s^{-1}}$ and amplitude $A=0.04 \,\mathrm{m}$. Then, sine oscillations with these characteristics were simply superimposed onto the simulated trajectories $\boldsymbol{z}(t)=x(t)+\boldsymbol{j} y(t)$, viz.
\begin{equation}
\boldsymbol{z}'(t)=\boldsymbol{z}(t)+A\,e^{\boldsymbol{j}\omega t}\boldsymbol{e_{\perp}}(t).
\end{equation}

\section*{Acknowledgments}
The authors are grateful to Chuan-Zhi (aka Thomas) XIE, Hakim BENKIRANE, and Iker ZURIGUEL for their input, 
to Hisashi MURAKAMI, Claudio FELICIANI, Mohcine CHRAIBI, Daniel PARISI, and Mehdi MOUSSAID for
so readily accepting to share their experimental data. This work was partly conducted in the frame of
the French-German research project MADRAS
funded in France by the Agence Nationale de la Recherche (grant number ANR-20-CE92-0033), and in
Germany by the Deutsche Forschungsgemeinschaft (grant number 446168800). I.E. acknowledges Ministerio de Economía y Competitividad (Spanish Government) through Project
No. PID2020-114839GB-I00 MINECO/AEI/FEDER, UE and Asociación de Amigos de la Universidad de Navarra for their economical support.

\clearpage

\setcounter{equation}{0}
\renewcommand\theequation{S\arabic{equation}}    
\setcounter{figure}{0}
\renewcommand\thefigure{S\arabic{figure}}    
\renewcommand{\theHfigure}{S\arabic{figure}}
\setcounter{section}{0}
\renewcommand{\thesection}{\Alph{section}} 

\section{Derivation of the bio-mechanical term energy}
\label{sub:app_SpeedEnergy}

The literature in physiology relates the energy expenditure of walking to the rate of oxygen consumption ($V_{O_2}$), which has a ``rest'' component and a speed-dependent component:

\begin{equation}
    V_{O_2} = V_{O_2}^{(rest)} + V_{O_2}^{(walking)}
\end{equation}

We are interested in the second contribution which, in the experimental work of Ludlow et al. \cite{ludlow2016energy} is reasonably well fitted by an equation of the form:

\begin{equation}
    e^{speed}[u] = K_{\mathrm{s1}} + K_{\mathrm{s2}}u^2, \, \text{for}\,  u \geq u_c
\end{equation}

This quadratic relation is consistent with other empirical studies where the energy expenditure of humans in walking motion has been also studied \cite{pandolf1977predicting}. For the particular case of this work, we discard the base energy consumption (i.e., $e^{speed}[0\ms]$) and subtract this contribution from the experimental data. After this process, we find that the coefficients of the previous equation must be such that:

\begin{equation}
    e^{speed}[1.5\ms] \approx  3\cdot e^{speed}[0.5\ms]
\end{equation}

Finally, we choose to smoothly connect the above $e^{speed}$ expression to 0 so as to avoid discontinuities. This is done with a second-order polynomial:

\begin{equation}
    e^{speed}[u] = K_{\mathrm{s3}}u + K_{\mathrm{s4}}u^2, \, \text{for}\,  u < u_c
\end{equation}

with coefficients such that they match the higher-speed curve at $u = u_c$, for the single-point value and the derivative. Taking the value of $u_c \simeq 0.1\ms$ and $e^{speed}[1\ms] = 1$, we are able to calculate the value of the 4 parameters associated with both equations, arriving at:

\begin{equation}
\begin{cases}
 K_{\scale{s1}} = 0.4\\
 K_{\scale{s2}} = 0.6 \\
 K_{\scale{s3}} = 7.6 \\
 K_{\scale{s4}} = -35.4 
\end{cases}
\end{equation}

so that

\begin{equation}
\displaystyle
    e^{\mathrm{speed}}(u)=
    \begin{cases}
7.6\,u-35.4\,u^{2} & \text{ for } u <0.1\ms\\
0.4+0.6\,u^{2} & \text{ for }u\geqslant 0.1\ms.
\end{cases}
\end{equation}

\begin{figure}[h]
\centering
\includegraphics[width=0.47\columnwidth]{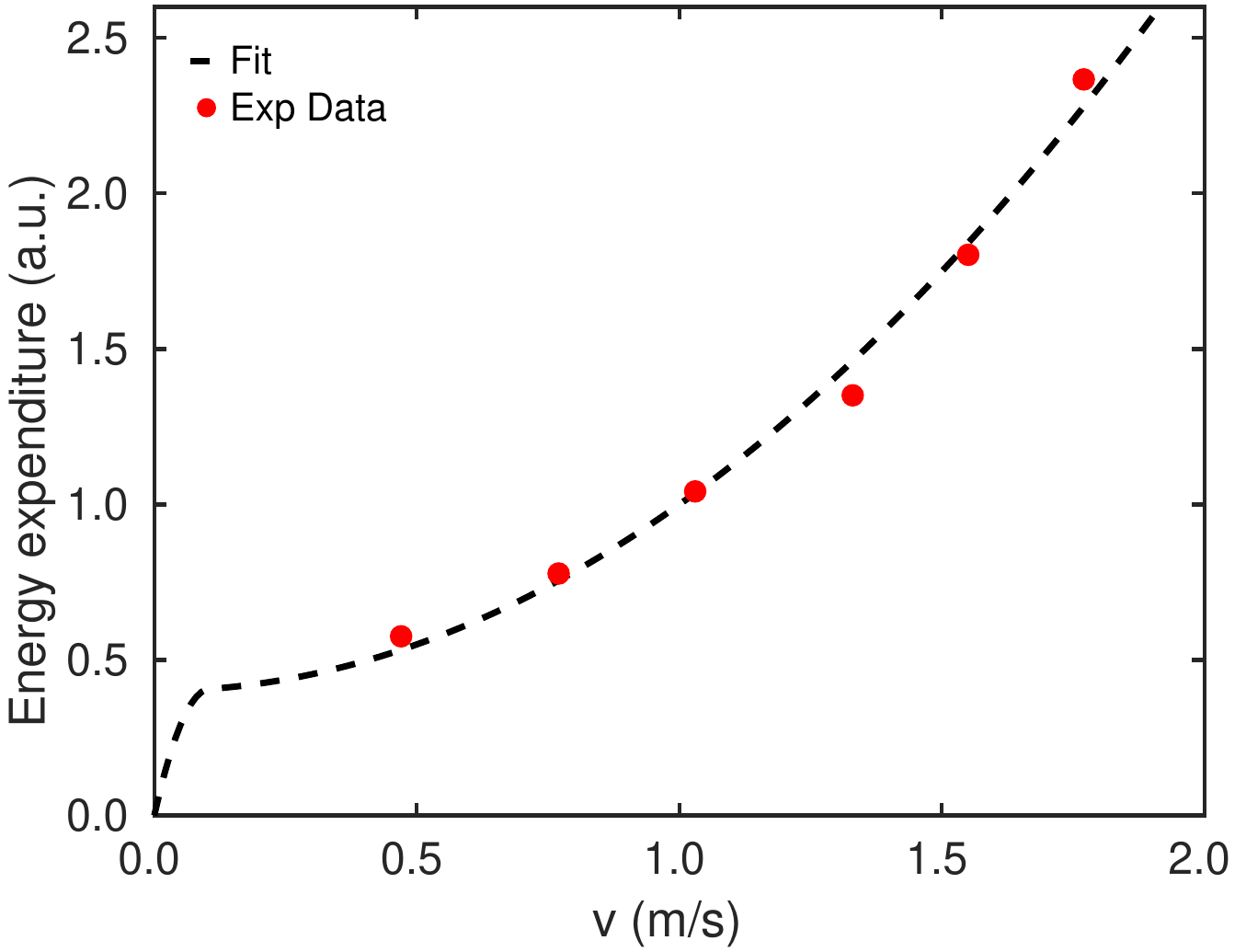}
\caption{ Bio-mechanical cost $e^{\mathrm{speed}}$ associated with the walking speed $v$ in the ANDA model, given by \eq{eq:Espeed}, and comparison to the aggregated data of \cite{ludlow2016energy} (adult group).}
\label{fig:Espeed}
\end{figure}

This functional form nicely fits the experimental data of  \cite{ludlow2016energy}, as shown in Fig.~\ref{fig:Espeed}.

\section{Derivation of the free walking speed}
\label{sub:app_Free}
To express the free-walking speed of agents as a function of the ANDA model parameters, let us consider isolated agents. By definition, they have no interactions with other agents or the built environment; the perceived cost for motion is thus 
$\E(\boldsymbol{u})= \delta t \Big[ e^{\mathrm{speed}}(\boldsymbol{u}) + e^{\mathrm{inertia}}(\boldsymbol{u}) \Big]+ \E^T(\boldsymbol{r}+\delta t\, \boldsymbol{u})$. Interestingly, this function (with its explicit dependences
given by \eq{eq:ET_D}, \eq{eq:Espeed} and \eq{eq:EInertia}) has the same qualitative dependence on
the (longitudinal) speed $u$ as the potential empirically estimated by Corbetta and co-workers from their tracking of dilute (i.e., non-interacting) pedestrians walking on a staircase landing (Fig.~5 of \cite{corbetta2017fluctuations}), with a local minimum at $u=0$ and a global minimum at the free walking speed $u^{\infty} \approx 1\ms$. 
\rev{In the stationary state, the actual walking speed matches the desired one, so that $\boldsymbol{u}^{\infty}$
is obtained by extremizing $\E(\boldsymbol{u})$,}
{Note however that, contrary to \cite{corbetta2017fluctuations}, no exogeneous noise is inserted into our decisional layer. Therefore, in the stationary state, the actual walking speed of a given agent does not fluctuate; instead, it  matches the desired one  $\boldsymbol{u}^{\infty}$, which is obtained by extremizing $\E(\boldsymbol{u})$, }
  viz.,

\begin{eqnarray}
0	&=&	\frac{1}{\delta t} \nabla_{\boldsymbol{u}}\E\Big|_{\boldsymbol{u}=\boldsymbol{u}^{\infty}} \\
	&=&	2\,\mu\Big(\boldsymbol{u}^{\infty}-\boldsymbol{v}(t)\Big)+\frac{de^{\mathrm{speed}}}{du}\frac{\boldsymbol{u}^{\infty}}{u^{\infty}} + \nabla_{\boldsymbol{r}}\E^{T} \\
	&=&	2\,\mu\Big(\boldsymbol{u}^{\infty}-\boldsymbol{v}(t)\Big)+1.2\,\boldsymbol{u}^{\infty}- K^{T}\,\boldsymbol{t}
	\end{eqnarray}
where we have used the expression of $e^{\mathrm{speed}}$ for $u\geqslant0.1\ms$ from \eq{eq:Espeed} and defined the unit vector  $\boldsymbol{t}=-\nabla_{ \boldsymbol{r}} \mathcal{D}/n(\boldsymbol{r})$ pointing towards the target.

Now, an isolated agent quickly reaches his/her desired velocity $\boldsymbol{u}^{\infty}$, so that the first term vanishes in the steady state. Therefore,
we arrive at $\boldsymbol{u}^{\infty}= \frac{K^T}{1.2} \boldsymbol{t}$, which can be used
to set the coefficient $K^T$ from the free-walking speed $u^{\infty}$.

\section{Avoidance of obstacle of non-convex shape}
\label{app:obstacle_other_models}

\begin{figure}[h]
\begin{center}
\includegraphics[width=0.75\columnwidth]{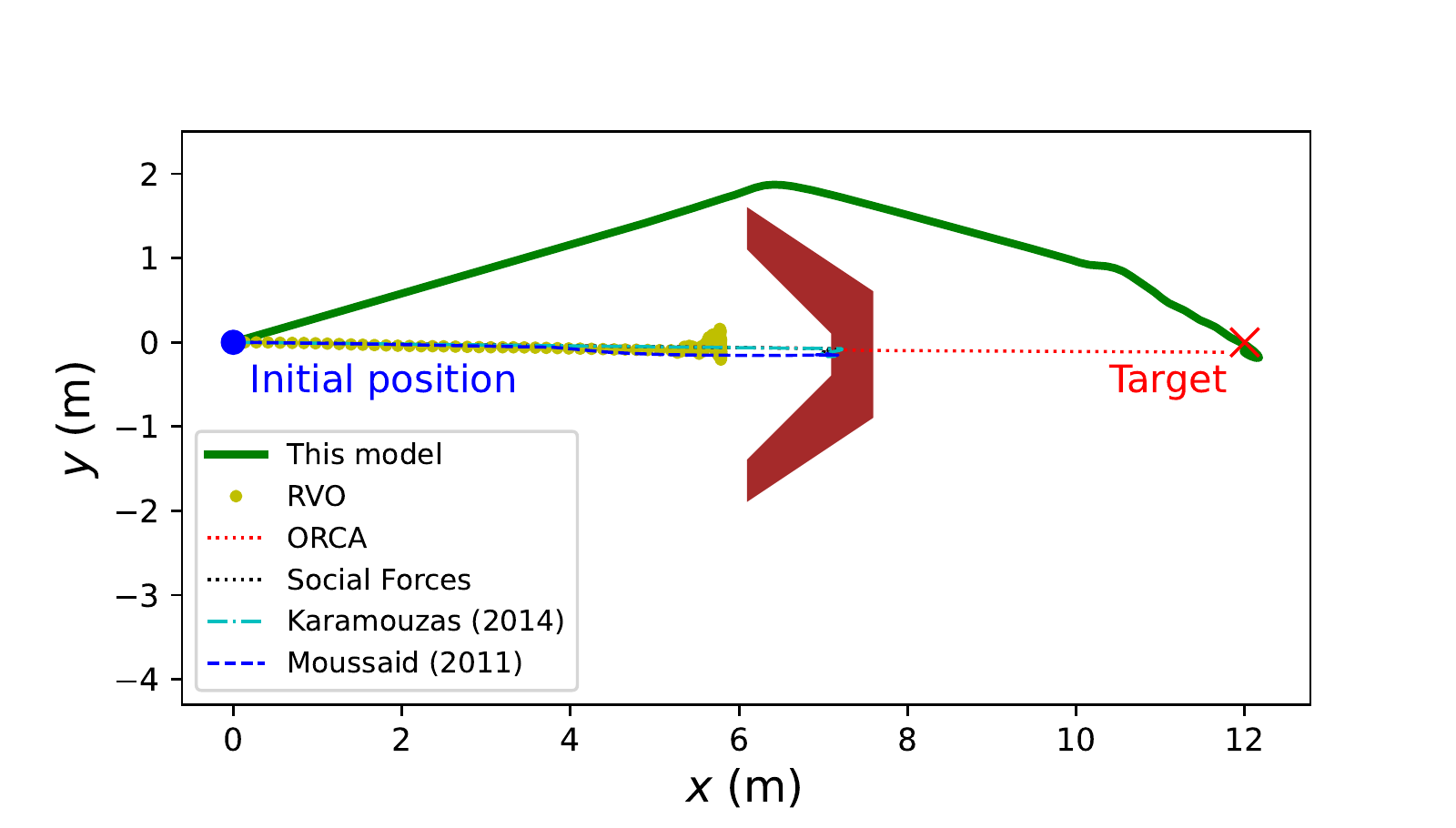}
\end{center}
\caption{Local navigation of one agent around a non-convex obstacle towards a predefined target, simulated with ANDA (`this model') and with alternative agent-based models: RVO \cite{van2010optimal}, ORCA \cite{van2011reciprocal}, social force model \cite{helbing1995social}, Karamouzas et al.'s TTC-based model \cite{Karamouzas2014universal}, Moussaid et al.'s heuristic model \cite{moussaid2011simple}. These other models were simulated using their implementation in the UMANS software with their native parameters in this software.}
\label{fig:SI_obstacle}
\end{figure}

The pedestrian's will to move is accounted by a floor field in ANDA. This provides a convenient handle
on the agents' motivation or haste to walk on uncomfortable ground or to stand too close to a wall (via the `refractive index' introduced in our Eikonal equation, in Sec.~\ref{sub:floor_field}), and more generally on their route choice. While one might argue that these elements belong to the tactical level, and not to the operational one, 
we contend that
\rev{the floor field also contributes to
improving the local navigation of the agents,}
{the recourse to a floor field considerably simplifies the way in which the local navigation of the agents is handled,}
notably around obstacles of arbitrary shape. To this end,
we consider a non-convex obstacle lying on an agent's path and compare in Fig.~\ref{fig:SI_obstacle} the output of ANDA and several very popular agent-based models, simulated with the UMANS
software developed at INRIA \cite{van2020generalized}. Manifestly,
for this test, only ANDA yields a reasonable result\rev{.}{, whereas the other models are in critical need of a pathfinding algorithm specifying how to circumnavigate the obstacle.}

\section{Consistency of the theoretical framework and differences with previous structures}
\label{app:eq_structure}

In the main text, we emphasized the importance of the sound delineation between the decision-making layer and the mechanical layer in ANDA. This delineation entails differences in the structure of the equations as compared to that of the Social Force Model \cite{helbing1995social} or Moussaid et al.'s heuristic model \cite{moussaid2011simple}; their implications are underscored here.

Schematically, instead of our \eq{eq:Newton_mech}, the former posits that 
\begin{equation}
m \boldsymbol{\ddot{r}}_j= m \frac{\boldsymbol{u}^{\infty}_j-\boldsymbol{u}_j}{\tau} + \boldsymbol{F}^{\mathrm{mech}}_{\rightarrow j} + \boldsymbol{F}^{\mathrm{soc}}_{\rightarrow j},
\label{eq:SI_SFM}
\end{equation}
where $\boldsymbol{F}^{\mathrm{mech}}_{\rightarrow j}$ and $\boldsymbol{F}^{\mathrm{soc}}_{\rightarrow j}$ refer 
to the mechanical and social forces exerted on $j$, respectively. 
\rev{}{It is true that, from Eq.~\ref{eq:SI_SFM}, the ANDA framework can be recovered if one sets $\boldsymbol{F}^{\mathrm{soc}}_{\rightarrow j}$ to $\frac{m}{\tau}(\boldsymbol{u}^{\star}_j-\boldsymbol{u}^{\infty}_j)$, but we will see below that in doing so the origin of the relaxational processes is misassigned.}

Moussaid et al.'s model \cite{moussaid2011simple} differs from our ANDA
framework in a more
subtle way, insofar as in both models the social
environment ($\boldsymbol{F}^{\mathrm{soc}}_{\rightarrow j}$ in \eq{eq:SI_SFM}) affects the choice of the desired velocity $\boldsymbol{u}^{\infty}_j$, instead of entering \eq{eq:SI_SFM}. 
\rev{
Still, the meaning of the characteristic time $\tau$ remains ambiguous, because it mingles a mechanical relaxation process with a decisional one (the heuristically determined desired velocity can change abruptly in that model).}
{But, in \cite{moussaid2011simple}, temporal variations in the heuristically determined desired velocity are not penalized, so that it can change very abruptly. These abrupt variations of the desired velocity are not immediately mirrored by the actual velocity, because they are damped by the timescale $\tau$ in Eq.~\ref{eq:SI_SFM}. The latter timescale is thus
 ambiguous, because it mingles a mechanical relaxation process with a decisional one. }
By contrast, ANDA penalizes sudden changes in the \emph{desired velocity} (via the term controlled by $\mu$ in the perceived cost and via the sequential update of $\boldsymbol{u}^{\infty}_j$ every $\delta t$) and then includes a \emph{mechanical relaxation} time governed by
$\tau^{\mathrm{mech}}$.

To \rev{illustrate this point with}{underline these differences using}  clear-cut examples, suppose that someone is walking on a moving walkway or a treadmill;
any variation of the speed of the apparatus will be transmitted to the pedestrian within a typical time $\tau^{\mathrm{mech}}$,
irrespective of the decisional layer (i.e., irrespective of $\delta t$ or $\mu$). Along the same vein, should one wish to describe a swimmer, the lower friction of the swimmer's body with the water (compared to the ground) will translate into a longer
mechanical relaxation time $\tau^{\mathrm{mech}}$. Conversely, the slower responses of distracted pedestrians (Sec.~\ref{sub:distracted}) or older people can readily be transcribed into the decisional layer of ANDA but have no impact on $\tau^{\mathrm{mech}}$. Our framework, therefore, clarifies the distinct relaxational processes that
were amalgamated in other models and misled some practitioners into ill-founded calibrations 
of some model parameters (whether it be relaxation times or the mass $m$ in \eq{eq:SI_SFM}).

Note that ANDA remains compatible with the framework developed by van Toll et al. \cite{van2020generalized}, who
recast a variety of microscopic models by defining a generalized velocity cost, provided that the generalized cost
function can include an `inertial' term penalizing sudden changes in velocity (which was not the case for the models
implemented so far in this framework).

\section{The intruder problem}
\label{sec:Intruder}

While competitive evacuations display many similarities with granular flows through a bottleneck \cite{Zuriguel2014clogging}, probably
owing to the prominence of mechanical contacts, recent experiments have shown that, surprisingly, the granular analogy fares much worse when a group
of \emph{static} people is crossed by an `intruder' \cite{nicolas2019mechanical}: Anticipation and self-propulsion by the pedestrians then play a major role in opening a pedestrian-free tunnel
ahead of the intruder via \emph{transverse} displacements, in stark contrast with \rev{the granular case}{the stationary response of a granular mono-layer (where \emph{non-transverse} recirculation eddies are observed); see the displacement fields in \cite{nicolas2019mechanical}}. This holds even in the dense regime, where mechanical forces were believed to prevail.

The Social Force Model goes completely amiss in the description of these features \cite{raj2021moving}, \rev{which in principle}{whereas in principle they} could be captured by our model: agents in the static crowd can anticipate a risk of collision with the intruder and move `out of harm's way' in advance, by walking away from the \rev{expected (linear) intruder's trajectory}{intruder's path}. 
\rev{In practice, however,}{Indeed,} we managed to reproduce the density field, with a `tunnel-like' opening ahead the intruder due to anticipation.
{\rev, but not}{On the other hand, we failed to replicate } the purely transverse displacements observed experimentally, even with slight variations of the model or its parameters.
In a parallel paper dedicated to this scenario \cite{bonnemain2022pedestrians}, we ascribed the deficiency of most existing agent-based models (including a variant of ANDA, see \emph{SI} of \cite{bonnemain2022pedestrians}) regarding this effect to
the fact that, in this situation, 
\rev{the local navigation is mingled with tactical planning }
{tactical planning interferes with the local navigation} and that the modeled agents are too short-sighted to achieve this anticipation. 
For ANDA, the alleged origin of this deficiency can be pinpointed more precisely thanks to its transparent derivation: Taking the limit $\delta t \to 0$ in the anticipated cost of motion in \Cref{eq:E_T} sweeps away the possibility to plan a move that involves a \emph{non-constant} velocity \rev{}{$\boldsymbol{u}(t')$}.

A natural way to recover it would be to perform the optimization of the full time-integral in \Cref{eq:E_T}, i.e. with respect to the planned velocity function $\boldsymbol{u}(t'),\ t'\geqslant t$, as in game theory, at the expense of an unbearable computational cost.
Mean-field game theory can overcome this intractability, at the expense of losing sight of the discrete nature of pedestrians; see \cite{bonnemain2022pedestrians}.

Interestingly, this also explains the aforementioned hesitancy of some pedestrians when crossing a group, in the complex scenario studied in Sec.~\ref{sub:complex_geometry}\rev{}{: ANDA agents are somewhat too short-sighted in their planning to cross static groups efficiently}.

\section{Density Field - A Misleading Indicator of Stop-and-Go Movement}

Stop-and-go waves traditionally mark the onset of instability in unidirectional traffic at high density.
For the dynamics of crowds in a corridor, we argue that the evolution of the (linear) density is a poor indicator for the detection of such waves, 
for at least two reasons. 

\begin{itemize}
    \item First, density is averaged across the corridor width, whereas the jammed phases do not necessarily span the whole corridor width, as we confirmed by direct visualization of the simulated flows (see \href{https://drive.google.com/file/d/1QiTdEIZapVgtxrpLH9ULeKshqqyc6_CA/view?usp=sharing}{Movie S4}); this issue becomes all the more problematic as the corridor is wide. It also comforts the idea \cite{jin2021pedestrian} that pedestrians in a wide corridor can sometimes evade jammed regions through transverse motion.

    \item Secondly, the difference between the density $\rho_j$ in jammed phases and the density $\rho_f$ in flowing ones is fairly small, since the TTC term gets people to brake ahead of a halted person. Because in a strictly one-dimensional setting the conservation of the number of agents imposes that $\rho_f\,(v_f + |w|)\simeq\rho_j\,|w|$ in the steady state, where $v_f$ is the pedestrian speed in the flowing phase and $w$ is the stop-and-go wave speed, the small difference between $\rho_f$ and $\rho_j$ entails a fairly large wave speed $|w|\simeq \frac{\rho_f \, v_f}{\rho_j - \rho_f}$. And, indeed, in \Cref{fig:Speed-density}(2), we measure a wave speed $|w|\approx 2\, m/s$ larger than the free walking speed. We should note that this value exceeds what is typically found for stop-and-go waves \emph{in single pedestrian files}, where $|w|$ generally lies below $1\, m/s$ \cite{portz2011analyzing}. 
\end{itemize}

\section{Exit Capacities}
In this section, we detail the quantitative results obtained in simulations of evacuation from a crowded rectangular room, expanding
on the discussion initiated in the main text.

The bottleneck flow at the door is probed by computing the specific capacity $J_s=\frac{1}{w}\cdot\frac{N}{T_N}$ (where $T_N$ is the duration it took to evacuate $N$ agents in the pseudo-stationary stage of the evacuation), i.e., mean flow rates per unit width of the door, for different door widths and agents' eagerness to escape, using $u^{\infty}$ as a proxy for the latter. 

For a $1\,\mathrm{m}$-wide door, our simulations yield $J_s= 1.80\,\mathrm{ped/m/s}$ in normal conditions ($u^{\infty}=1.5\ms$), right between the estimate $J_s=1.60\,\mathrm{ped/m/s}$ reported in \cite{predtechenskii1978planning} and the experimental
measurements $J_s=1.85\,\mathrm{ped/m/s}$ and $J_s=1.90\,\mathrm{ped/m/s}$ in \cite{kretz2006experimental} and in \cite{seyfried2009new}, respectively. In these last two publications, the specific capacity decreases slightly to around $J_s=1.8\,\mathrm{ped/m/s}$ and $J_s=1.6-1.7\,\mathrm{ped/m/s}$, respectively, when the door is narrowed to $w=80\,\mathrm{cm}$; it drops somewhat more significantly in ANDA, to $J_s=1.54\,\mathrm{ped/m/s}$, still for  $u^{\infty}=1.5\,\mathrm{m/s}$, but the agreement remains acceptable.

For \emph{even} narrower doors ($w=70\,\mathrm{cm}$) under \emph{competitive} settings ($u^{\infty}\simeq 3.0\ms$) it must plainly be conceded that a marked discrepancy arises in the absolute values of the specific capacity, which is around $J_s\approx 1.4\,\mathrm{ped/m/s}$ in the simulations and around $J_s\approx 3.6\,\mathrm{ped/m/s}$ in the experiments.
This can easily be explained: Our approximation of pedestrians as frictionless disks is stretched beyond any reasonable limit in a regime dominated by mechanical obstructions and contacts, and in which the shape of agents matters considerably \cite{echeverria2020pedestrian}. 
A better physical description would be attained by refining the mechanical layer using more realistic agent shapes.

\section{Additional figures}
\clearpage

\begin{figure}
\includegraphics[width=0.95\textwidth]{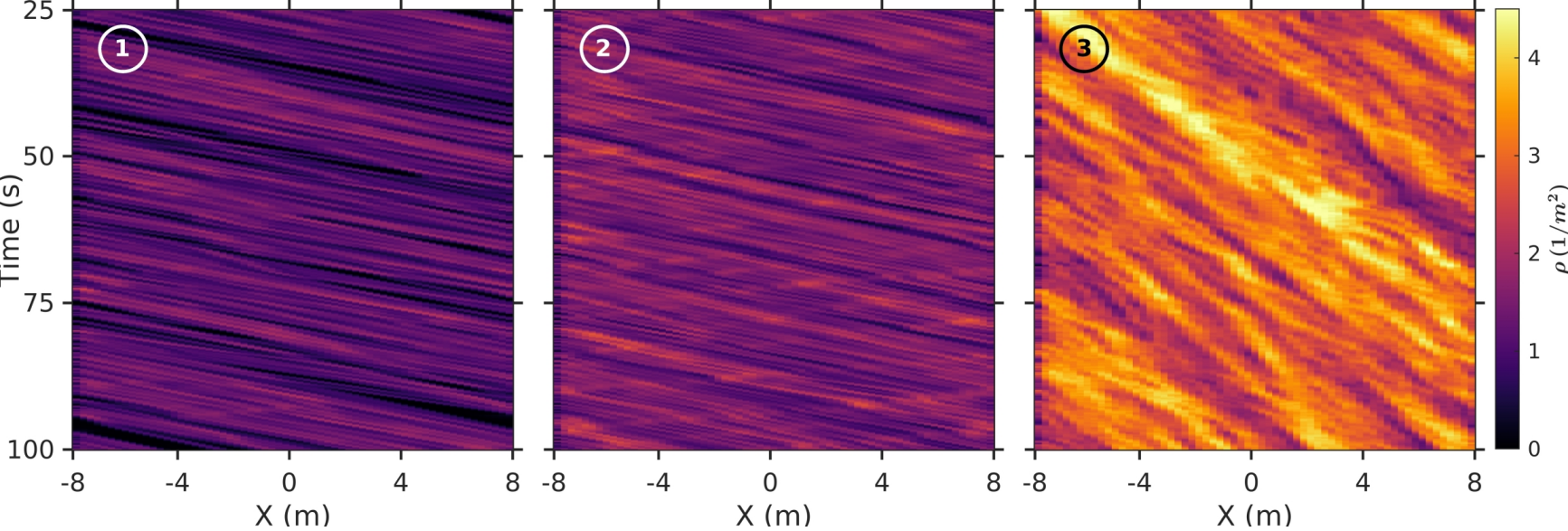}
\caption{Spatio-temporal diagrams of the coarse-grained local density, represented at different global densities as indicated in the titles of each panel. The data employed in their construction is identical to the data displayed in the velocity fields of \Cref{fig:Speed-density}.}
\label{fig:SI_DensityMaps}
\end{figure}

\clearpage

\begin{figure}
\includegraphics[width=0.95\textwidth]{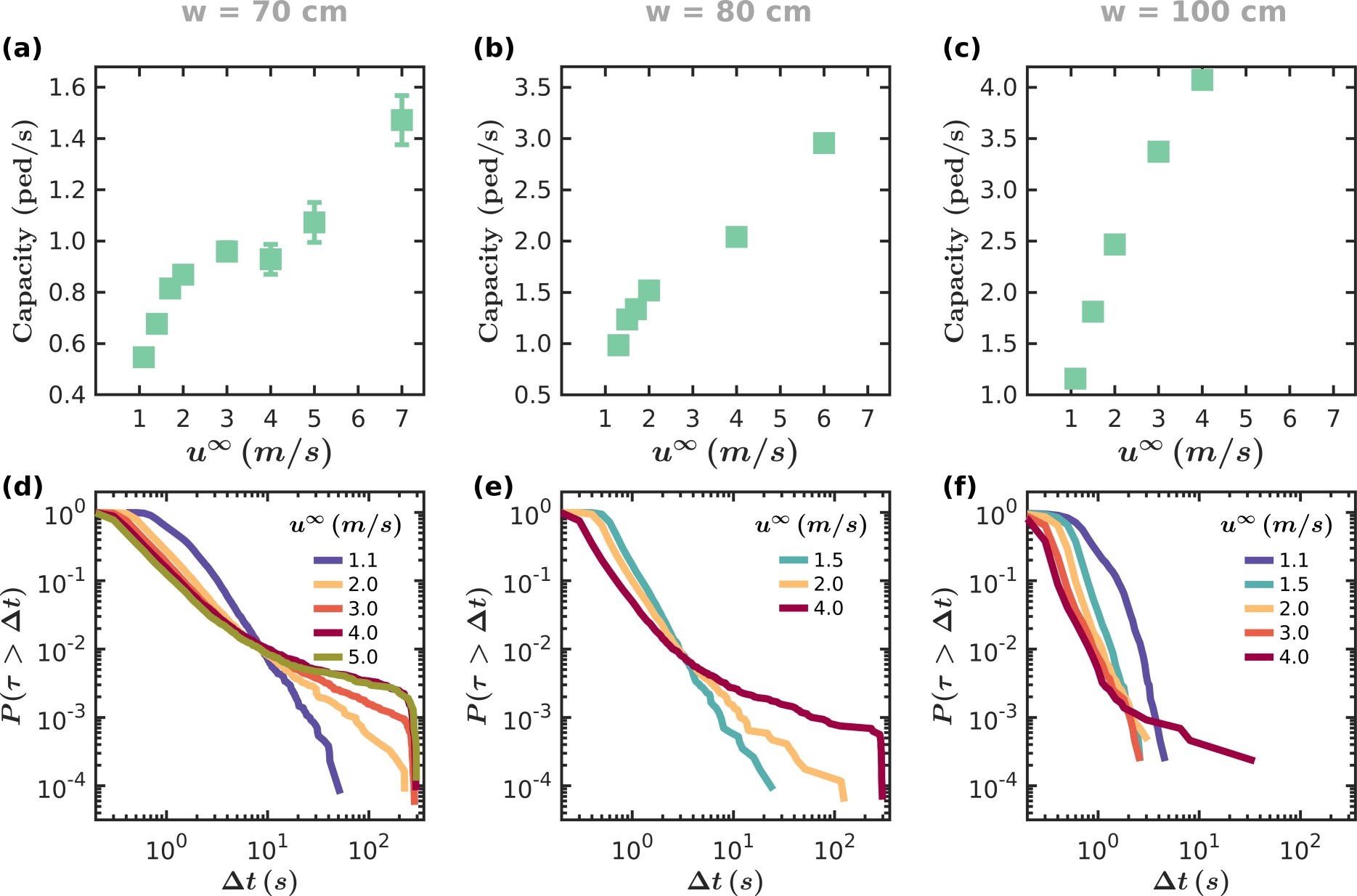}
\caption{Evacuation dynamics through bottleneck of different widths, {\bf (a,d)} $w=70\,\mathrm{cm}$,  {\bf (b,e)} $w=80\,\mathrm{cm}$,  {\bf (c,f)} $w=100\,\mathrm{cm}$. The top row shows the exit capacity as a function of the preferential speed $u^{\infty}$; the bottom row exposes the survival functions $P(\tau>\Delta t)$ of time gaps $\tau$ between successive egresses.}
\label{fig:SI_evac_Multidoor}
\end{figure}

\clearpage

\begin{figure}
\includegraphics[width=0.95\textwidth]{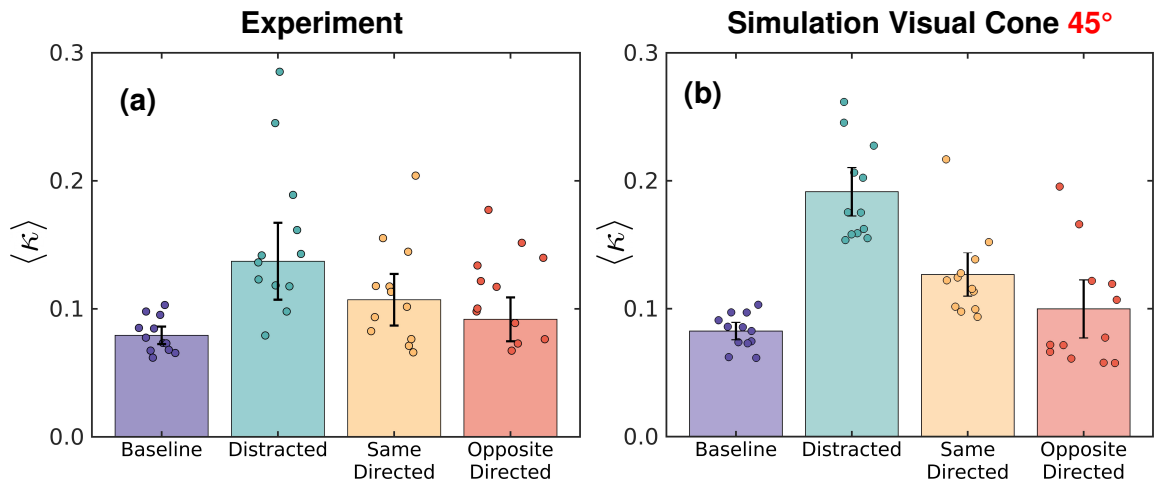}
\caption{{\bf(a)} Suddenness of turn $\kappa$ calculated for the experiment of Murakami et al. \cite{murakami2021mutual} already presented in \Cref{fig:Murakami}, main text; {\bf(b)} Suddenness of turn $\kappa$ computed in our model reducing agents' half-angle of the visual to $\theta=45^{\circ}$.}
\label{fig:SI_Murakami45}
\end{figure}

\newpage

\blue{
\begin{figure}
\includegraphics[width=\textwidth]{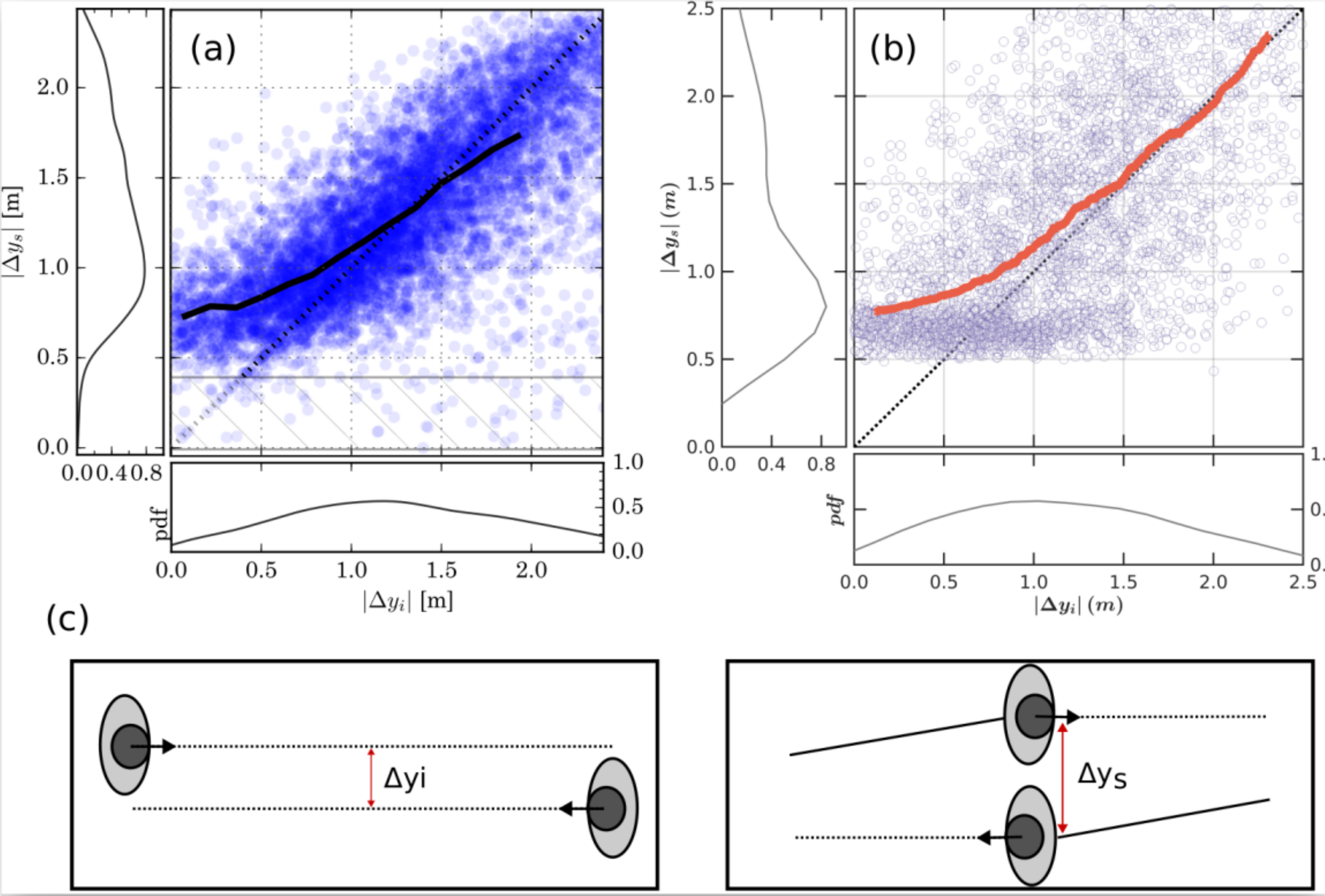}
\caption{Dependence of the absolute lateral distance at the beginning of the avoidance maneuver [$|\Delta y_i|$, x-axis; see left panel (c)] on the absolute lateral distances when passing side-by-side [$|\Delta y_s|$, y-axis; see right panel (c)]. Each data point in the scatter plot corresponds to a pair of counter-walking pedestrians: (a) Empirical data from \cite{corbetta2018physics} (reproduced with permission); (b) Simulations of the ANDA model. Solid lines in both cases represent the ensemble-averaged value of $|\Delta y_s|$ conditioned on $|\Delta y_i|$. The dotted diagonal line $|\Delta y_s| = |\Delta y_i|$ (no transverse deviation), in which both pedestrians keep moving straight ahead, is shown for reference. The bottom and left panels in (a) and (b) display the probability density
functions (pdfs) of $|\Delta y_i|$ and $|\Delta y_s|$.
}
\label{fig:AvoidCorbetta}
\end{figure}
}
\clearpage

\begin{table}[]
\caption{Summary of empirical and experimental evidence in support of ANDA.}
\centering
\begin{tabular}{@{}lll@{}}
\toprule
\multicolumn{1}{c}{\textbf{Description}}      & \textbf{Type}                  & \multicolumn{1}{c}{\textbf{Reference}} \\ \midrule
Avoidance maneuvers of individual pedestrians & Empirical \& Experimental Data              & \cite{Moussaid2009experimental, corbetta2018physics}              \\
Intruder                                      & Experimental Data              & \cite{nicolas2019mechanical,bonnemain2022pedestrians}              \\
Speed density relation - Unidirectional Flow  & Empirical \& Experimental Data & \cite{Older1968,Mori,Weidmann1993,zhang2011transitions}              \\
Speed density relation - Bidirectional Flow   & Experimental Data              & \cite{zhang2012ordering}              \\
Lane Formation                                & Experimental Data              & \cite{jin2019observational}              \\
Bottleneck Flow                               & Experimental Data              & \cite{predtechenskii1978planning,kretz2006experimental,seyfried2009new}              \\
Phone Distraction                             & Experimental Data               & \cite{murakami2022spontaneous}              \\ \bottomrule
\end{tabular}
\label{tab:Scenarios}
\end{table}

\section*{Supplementary Movies}
\href{https://drive.google.com/file/d/1L2fGup_izpplfIDFQh0NeULRVmUrm2uK/view?usp=sharing}{(Movie 1)}. Simulated collision avoidance dynamics. The movie is made of two parts: (1) the avoidance maneuver of a moving pedestrian coming across a static pedestrian and (2) the avoidance maneuver of two counter-walking pedestrians in a head-on collision. Besides, the video includes in the upper part the energy maps originated by each pedestrian based on their perceived cost, which they will have to minimize for the election of a new velocity vector (white arrow). \url{https://drive.google.com/file/d/1L2fGup_izpplfIDFQh0NeULRVmUrm2uK/view?usp=sharing}

\href{https://drive.google.com/file/d/1XZL9ozB49iPGLCvTdyQIYNUbDNEucS3w/view?usp=sharing}{(Movie 2)}. Antipodal simulation with 10 regularly spaced pedestrians with identical preferential speeds. \url{https://drive.google.com/file/d/1XZL9ozB49iPGLCvTdyQIYNUbDNEucS3w/view?usp=sharing}

\href{https://drive.google.com/file/d/1fy2morxISAN6OxCAYTgxBmD8eIgvy0dg/view?usp=sharing}{(Movie 3)}. Navigation in a complex geometries. Simulation of 100 pedestrians within a geometry inspired by the ground floor of Montparnasse train station in Paris, France. Agents are colored according to their specific target zone. Thus, at the end of the movie, we can observe that pedestrians are grouped by color, showing that they have reached their destination. \url{https://drive.google.com/file/d/1fy2morxISAN6OxCAYTgxBmD8eIgvy0dg/view?usp=sharing}

\href{https://drive.google.com/file/d/1QiTdEIZapVgtxrpLH9ULeKshqqyc6_CA/view?usp=sharing}{(Movie 4)}. Simulated unidirectional flow in a 3m-wide corridor. The movie shows the performance of our numerical model for three different global densities $\mathbf{(\boldsymbol{\rho} = \lbrace 1,2,3 \rbrace ~ped/m^2)}$. Agents are colored depending on their current walking speed, according to the colobar on the right. \url{https://drive.google.com/file/d/1QiTdEIZapVgtxrpLH9ULeKshqqyc6_CA/view?usp=sharing}



\end{document}